\documentclass[twocolumn,trackchanges,floatfix]{aastex62} 

\newcommand{\bef}{\begin{figure}}      
\newcommand{\eef}{\end{figure}}      
\newcommand{\bea}{\begin{eqnarray}}    
\newcommand{\eea}{\end{eqnarray}}      
\newcommand{\be}{\begin{equation}}      
\newcommand{\ee}{\end{equation}}  

\usepackage{lineno}
\graphicspath{{./}{figures/}}

\received{???}
\revised{???}
\accepted{???}
\shorttitle{GAIA DR3 extended kinematic maps}
\shortauthors{Wang et al.}
\usepackage{amsmath}

\begin{document}

\title{Mapping the Milky Way Disk with Gaia DR3: 
3D extended kinematic maps and rotation curve to $\approx 30$ kpc}
\affil{Centro  Ricerche Enrico Fermi, Via Pansiperna 89a, 00184 Rome, Italy}
\author[0000-0001-8459-1036]{Hai-Feng Wang}
\affil{Centro  Ricerche Enrico Fermi, Via Pansiperna 89a, 00184 Rome, Italy}
\author{\v{Z}ofia Chrob\'akov\'a}
\affil{Faculty of Mathematics, Physics, and Informatics, Comenius University, Mlynsk\'a dolina, 842 48 Bratislava, Slovakia}
\author{Mart\'in L\'opez-Corredoira}
\affil{Instituto de Astrof\'\i sica de Canarias, E-38205 La Laguna, Tenerife, Spain}
\affil{Departamento de Astrof\'\i sica, Universidad de La Laguna, E-38206 La Laguna, Tenerife, Spain}
\author[0000-0003-2236-4537]{Francesco  Sylos Labini}
\affil{Centro  Ricerche Enrico Fermi, Via Pansiperna 89a, 00184 Rome, Italy}
\affil{Istituto Nazionale Fisica Nucleare, Unit\`a Roma 1, Dipartimento di Fisica, Universit\'a di Roma ``Sapienza'', 00185 Rome, Italy}

\correspondingauthor{HFW}
\email{haifeng.wang@cref.it};\\

\begin{abstract}

We apply a statistical deconvolution of the parallax errors based on 
Lucy's inversion method (LIM) to the Gaia-DR3 sources 
to measure their three dimensional velocity components in 
the range of Galactocentric distances $R$ between 
8 kpc and 30 kpc with their corresponding errors and root mean square values.
We find results that 
are consistent with those obtained by applying LIM to  the Gaia-DR2 sources, and 
we conclude that the method  gives convergent and more accurate results 
by improving the statistics of the data-set and lowering observational errors.
The kinematic maps reconstructed with LIM up to 
$R \approx 30$ kpc show that
the Milky Way is characterized by 
asymmetrical motions with significant gradients in all velocity components. 
Furthermore, we determine the Galaxy rotation curve $V_C(R)$ up 
to $\approx 27.5$ kpc with the cylindrical
Jeans equation assuming an axisymmetric gravitational potential. 
We find 
that $V_C(R)$ is significantly declining up to the largest
radius investigated. 
Finally, we also measure $V_C(R)$  
at different vertical heights, showing that, for 
$R <15$ kpc, there is a marked dependence on $Z$, 
whereas at larger $R$ the dependence on $Z$ is negligible.

\end{abstract}


\section{Introduction}

The Gaia mission \citep{Gaia_2016} is providing the most detailed Milky Way survey to date by measuring 
stellar astrometry, photometry, and spectroscopy. From  this data-set it is possible to  
derive the spatial distribution, kinematics, and many other physical properties 
of the Milky Way. The third Gaia data release (DR3) \citep{Gaia_DR3} contains  the same photometric magnitude and astrometric information
of the 
previous Gaia data EDR3 releases but for a wider sample of stars with new determinations of 
spectra, radial velocity, chemical abundance, value-added catalogs, etc. 
  
Concerning the kinematics of our Galaxy the Gaia data-sets provide the full astrometric solution - positions on the sky, parallax, and proper motion. Gaia DR3 has also provided a significant increase  of  the  stars line-of-sight velocity catalog, from 7,209,831 in DR2 to 33,812,183 in DR3 \citep{Katz_etal_2022}.

Since the first data release, Gaia has played a major part in revealing the kinematics of our Galaxy. \cite{Antoja2016,2017antoja} used Gaia DR1 and mock data to probe the influence of the spiral arms on Galactic kinematics, found that the typical difference in transverse velocity at symmetric longitudes is about 2 km s$^{-1}$, but can be larger than 10 km s$^{-1}$ at some longitudes and distances.

By using the DR2, \cite{Katz_etal_2018} revealed streaming motions in all three velocity components and found that the vertical velocities show a superposition of modes. The same work found small amplitude perturbations in the velocity dispersion as well.

\cite{Kawata_etal_2018} 
have  carried out a kinematic analysis for a radii $R<13$ kpc, 
a range in which the relative error in the distance was lower than 20 \%. \cite{Poggio_etal_2018} extended the analyzed region up to $R <15$  kpc in combination with 2MASS photometry, with the aim of characterizing vertical motions. By using around 5 million stars from the Gaia DR2 catalog belonging to the Milky Way disk, \cite{Ramos2018} have also detected many kinematic asymmetries whose azimuthal velocity decreased with the galactic radius.

\citet{Lopez-Corredoira_Sylos-Labini_2019} have presented extended kinematic maps of the Galaxy with Gaia DR2 including the region where the relative error in distance was between 20$-$100\%.
That was possible  thanks to the use of a statistical deconvolution algorithm of the parallax errors named
Lucy's inversion method (LIM) ( \citet{1974AJ.....79..745L} and reference therein). 

By applying this method to the Gaia-DR2 data-set,  and 
 including line-of-sight velocity measurements, 
 \cite{Lopez-Corredoira_Sylos-Labini_2019} 
 have extended the range of distances  for the kinematic analyses
of $\approx 7$ kpc with respect to those presented by \cite{Katz_etal_2018}, thus adding the range of Galactocentric distances between 13 kpc and 20 kpc to the previous maps. They
found velocity gradients of about 40 km s$^{-1}$ in the radial and azimuthal directions, 10 km s$^{-1}$ in the vertical direction, as well as north-south asymmetries. 

\cite{Bennett2019} made detailed measurements of the wave-like north-south asymmetry in the vertical stellar counts, which showed some deficits at heights $\approx$ 0.4, $\approx$ 0.9 and $\approx$ 1.5 kpc, and peaks at $\approx$ 0.2, $\approx$ 0.7 and $\approx$ 1.1 kpc. The mean vertical velocity is also found to show a north-south symmetric dip at $\approx$ 0.5 kpc with an amplitude of $\approx$ 2 km s$^{-1}$. By using OB stellar samples and RGB samples, \cite{Gomez2019} found that the median vertical proper motion values show a clear vertical modulation towards the anti-centre.

\cite{Antoja2021} used Gaia EDR3 to reveal that oscillations in median rotational and vertical velocities vary with radius, disk stars having large angular momentum moving vertically upwards and disk stars having slightly lower angular momentum moving preferentially downwards. 

\cite{Drimmel2022} traced the spiral structure and kinematics  using the young stellar population from the
Gaia DR3 data-set and found that the local arms are at least 8 kpc long, the outer arms are consistent with those seen in the HI survey. In addition, by analyzing the RGB samples 
with measured velocities, they found the streaming motions in the outer disk that may be associated 
to the spiral arms or bar dynamics.

Our aim in this paper is  to derive kinematic maps of the Galaxy
up to $R \approx 30$ kpc by applying LIM 
 on the Gaia-DR3 data-set only from the observational points of view. 
There are two advantages passing from DR2 to DR3: 
on the one hand the sub-sample including radial velocities extracted from Gaia-DR3 has a fainter magnitude limit, i.e.,
$G_{RVS}$ = 14 \citep{Katz_etal_2022}, 
than that from Gaia-DR2 ($G_{RVS}$ = 12) 
and thus many more sources; on the other hand  
parallax errors become lower passing from DR2 to DR3.
Indeed, according to \cite{Lindegren_etal_2021} 
uncertainties are smaller in DR3 with respect to DR2
by a factor 0.80 for 
parallaxes and positions and by a factor 0.51 for 
proper motions.

Moreover, by assuming that the Galaxy is in a steady state and that the gravitational potential is axisymmetric, i.e., by neglecting perturbations, we then can determine, by means of the Jeans equation, the velocity rotation curve $V_C(R)$ or $V_\phi$ up to $\approx 30$ kpc.
This allows us 
 to confirm 
the recent results by \cite{Eilers_etal_2019} 
 who found that the  $V_C(R)$  is 
 linearly  declining up to 25 kpc 
 with a slope of $\approx -1.7\pm 0.1$ $\mbox{km s}^{-1} \mbox{kpc}^{-1}$ (systematic uncertainty of 0.46 $\mbox{km s}^{-1} \mbox{kpc}^{-1}$). 
In addition, we  present estimations of the rotation curve at different heights and we compare them with the
results by \cite{Chrobakova_etal_2020}:  
most notably, we find, at small radii,  a marked dependence of $V_C(R)$ on $Z$ which will be shown below.

The paper is structured as follows: in Sect. \ref{sect2}, we briefly describe the Gaia DR3 sample 
we use in this work. Then in Sect. \ref{sect3} we outline essential elements of the LIM, while in Sect. \ref{sect4} 
we present results about the 3D kinematics and the rotation curve. Finally in Sect. \ref{sect5} we draw our main conclusions. 


\section{Gaia-DR3 data}
\label{sect2}

Gaia DR3  \cite{Gaia_DR3} provides line-of-sight velocities of more than 33 millions sources with a limiting magnitude of  $G \approx$ 14. The full astrometric solution has been done as 5-parameter fit for 585 million sources and 6-parameter fit for 882 million sources. While the median parallax uncertainty is 0.01$-$0.02 mas for $G$ \textless 15, 0.05 mas at $G =$ 17, 0.4 mas at $G =$ 20
and 1.0 mas at  $G = 21$.  The uncertainty in the determination of the proper motion is 0.02$-$0.03 mas yr$^{-1}$ for stars with $G$ \textless  15, 0.07 mas yr$^{-1}$ for stars with $G = 17$, 0.5 mas yr$^{-1}$ for stars with $G = 20$, 1.4 mas yr$^{-1}$ for stars with $G = 21$ {\citep{Lindegren_etal_2021}}.

Gaia DR3 also contains a release of magnitudes estimated
 from the integration of Radial Velocity Spectrometer (RVS) spectra for a sample of about 32.2 million stars brighter than $G_{RVS} \approx 14$
 (or $G \approx 15$). \cite{Sartoretti_etal_2022} has described the data used and the approach adopted to derive and validate the $G_{RVS}$ magnitudes published in DR3. They also provide estimates of the 
  $G_{RVS}$ pass-band and associated $G_{RVS}$ zero-point. \cite{Recio-Blanco_etal_2022} has summarized the stellar parametrization of the Gaia RVS spectra (around 5 millions) performed by the GSP-Spec module and published it as part of Gaia DR3. 
The formal median precision of radial velocities is 1.3 km s$^{-1}$ at $G_{RVS} = 12$   and 6.4 km s$^{-1}$ at $G_{RVS} = 14$. The velocities zero-point exhibit a small systematic trend with magnitude starting around $G_{RVS} = 11$  and reaching about 400 m s$^{-1}$ at $G_{RVS}  = 14$: to take into account this trend a correction formula is provided. Note the Gaia DR3 velocity scale is in satisfactory agreement with APOGEE, 
GALAH, GES and RAVE \citep{Katz_etal_2022}.

The Gaia-DR3 data-set provides for each 
star  the parallax $\pi$, the Galactic coordinates $(\ell,b)$, 
 the line-of-sight velocity $V_r$  and the two proper motions in equatorial coordinates 
 $\mu_\alpha \cos\delta$ and 
 $\mu_\delta$.
These six variables allow to place each star
in the 6D phase space, 
i.e., 3D spatial coordinates plus 3D
velocity coordinates.
The Galactocentric position in cylindrical coordinates has components: 
$R$, the Galactocentric distance,
$\phi$,  the Galactocentric azimuth 
and 
$Z$, the vertical distance. 
The Galactocentric velocity in cylindrical coordinates
has components:
 $V_R$, the radial velocity, 
$V_\phi$, the azimuthal velocity,
and 
$V_Z$, the vertical velocity. 
Details to transform 
respectively $(\pi,\ell,b)$ in 
Galactocentric  cartesian $(X,Y,Z)$ or  cylindrical $(R, \phi, z)$  coordinates
and 
$(V_r, \mu_\alpha \cos\delta, \mu_\delta)$ in $(V_R, V_\phi, V_Z)$
 are given in \cite{Lopez-Corredoira_Sylos-Labini_2019}. 
The 3D velocity we used is obtained by assuming the location of the Sun is R$_\odot$ = 8.34 kpc \citep{Reid_etal_2014} and Z$_\odot$ = 27 pc  \citep{Chen_etal_2001}. 
We use the solar motion values [U$_\odot$, V$_\odot$, W$_\odot$] = [11.1,12.24,7.25] km s$^{-1}$ \citep{Reid_etal_2014}. 
The value of the circular speed of the LSR is 238 km s$^{-1}$ \citep{Schonrich_etal_2010}. 
 Note that different solar values have negligible effect for this work.


\section{Lucy Inversion Method (LIM)}
\label{sect3}

It is known that Gaia parallax uncertainties increase with distance, in particular beyond 5 kpc from the Sun, and that they are also dependent on the star's magnitude. The large dispersion of parallaxes correspond to a  large dispersion of stellar distances: such dispersion affects  the value of mean distance at large heliocentric distances, which is indeed overestimated in a direct measurement with respect to the real one. 
In order to solve the problem of the deconvolution of the Gaussian errors with large root mean square (rms) 
values and the asymmetric parallax uncertainties, we adopt the LIM. We refer for more details to \citet{Lopez-Corredoira_Sylos-Labini_2019} where Monte Carlo simulations to test such  technique are also presented. 

The basic idea of the method is simple: in order to explore Galactic regions 
where relative parallax error are larger than $20\%$, i.e., for $R>$  18 kpc,
we can reduce the error in the distance determination by making averages over many stars.
We thus divide the observed Galactic 
region into $N_{cells}$ each containing many stars. We want 
to determine the average value 
of the velocity components and their dispersion
in each cell. The LIM provides the deconvolved  distribution of sources along the line-of-sight;
for this quantity we can estimate 
the mean heliocentric distance of all the stars included in a bin centered at  parallax $\pi $ and the corresponding variance $\sigma _{\overline{r}}^2(\pi )$. 
It is worth noticing that the LIM is model independent: we can recover information on the stellar distribution without introducing any prior.
In addition the LIM does not provide the distance
for each star, rather we only have statistical 
determination of the distance of a three-dimensional cell that typically contains many stars,
i.e., in each of the $N_{cell}$  in which we have divided the observed Galactic region.
In this way, in each cell 
we have the estimation of the average values of the distance, velocity components and their corresponding errors.


\section{Results}
\label{sect4}
\subsection{Velocity asymmetries with DR3}


\begin{figure*}
 \centering
 \includegraphics[width=0.9\textwidth]{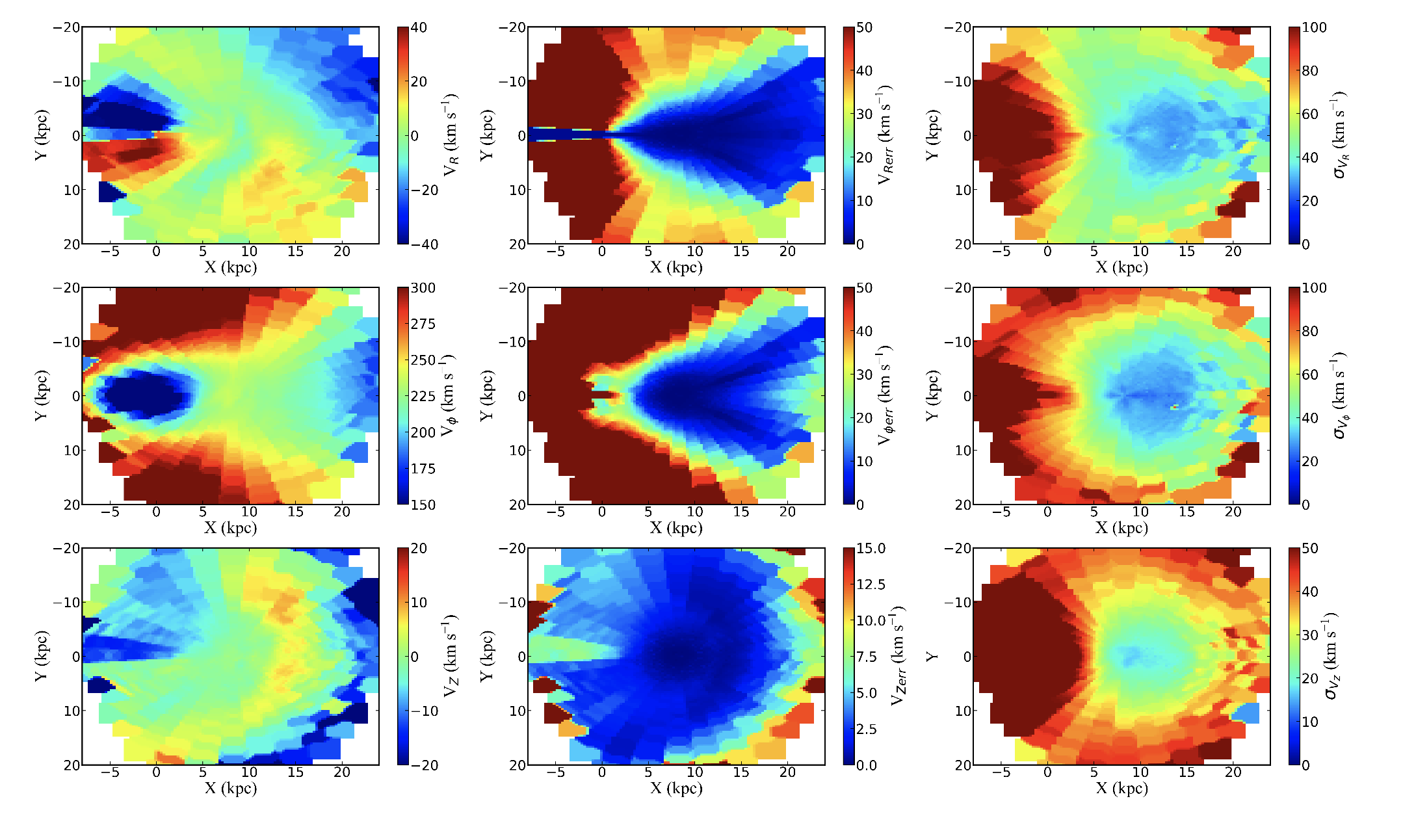}
 \caption{ %
For each velocity component (from top to bottom, respectively, 
$V_R \,, V_\phi\,,V_Z$)  we plot the map 
reconstructed by the LIM and  projected 
onto the Galactic  $(X,Y)$ plane
  of 
 the  average 
velocity  component  in km/s (left panels),
errors in km/s   (middle panels)
and 
rms value in km/s
corrected
for measurement errors
(right panels).
  } 
  \label{lucy2}
\end{figure*}
\begin{figure*}
  \centering
  \includegraphics[width=0.9\textwidth]{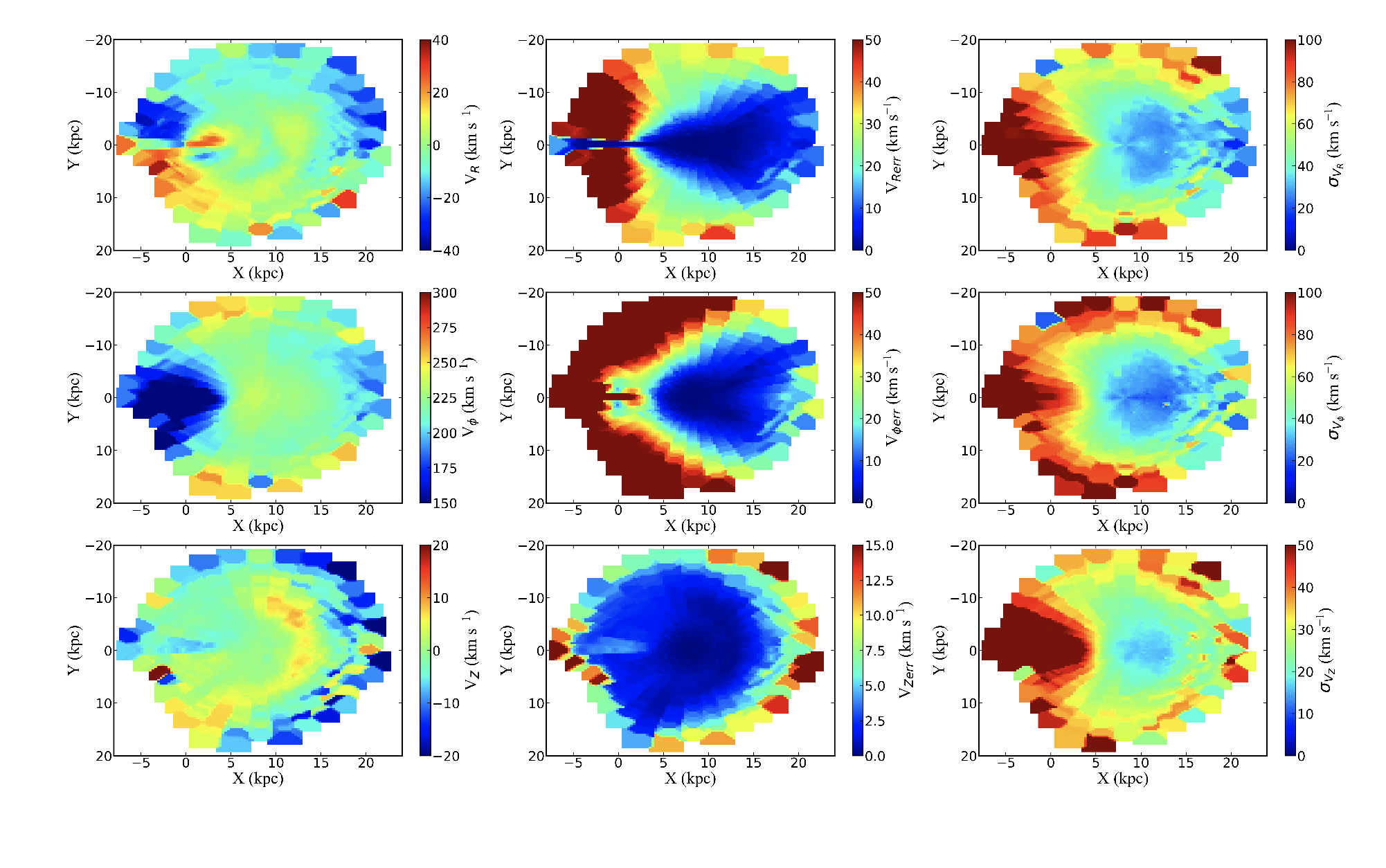}
   \caption{As Fig.\ref{lucy2} but including the correction due to the zero-point bias. 
Note that the zero-point correction particularly affects 
 large radii by introducing a larger noise/signal ratio: for this
 reason such correction reduces the range of radii where we can reconstruct 
 kinematic properties.
}  
\label{lucy2zp}
\end{figure*}
\begin{figure}
  \includegraphics[width=0.45\textwidth]{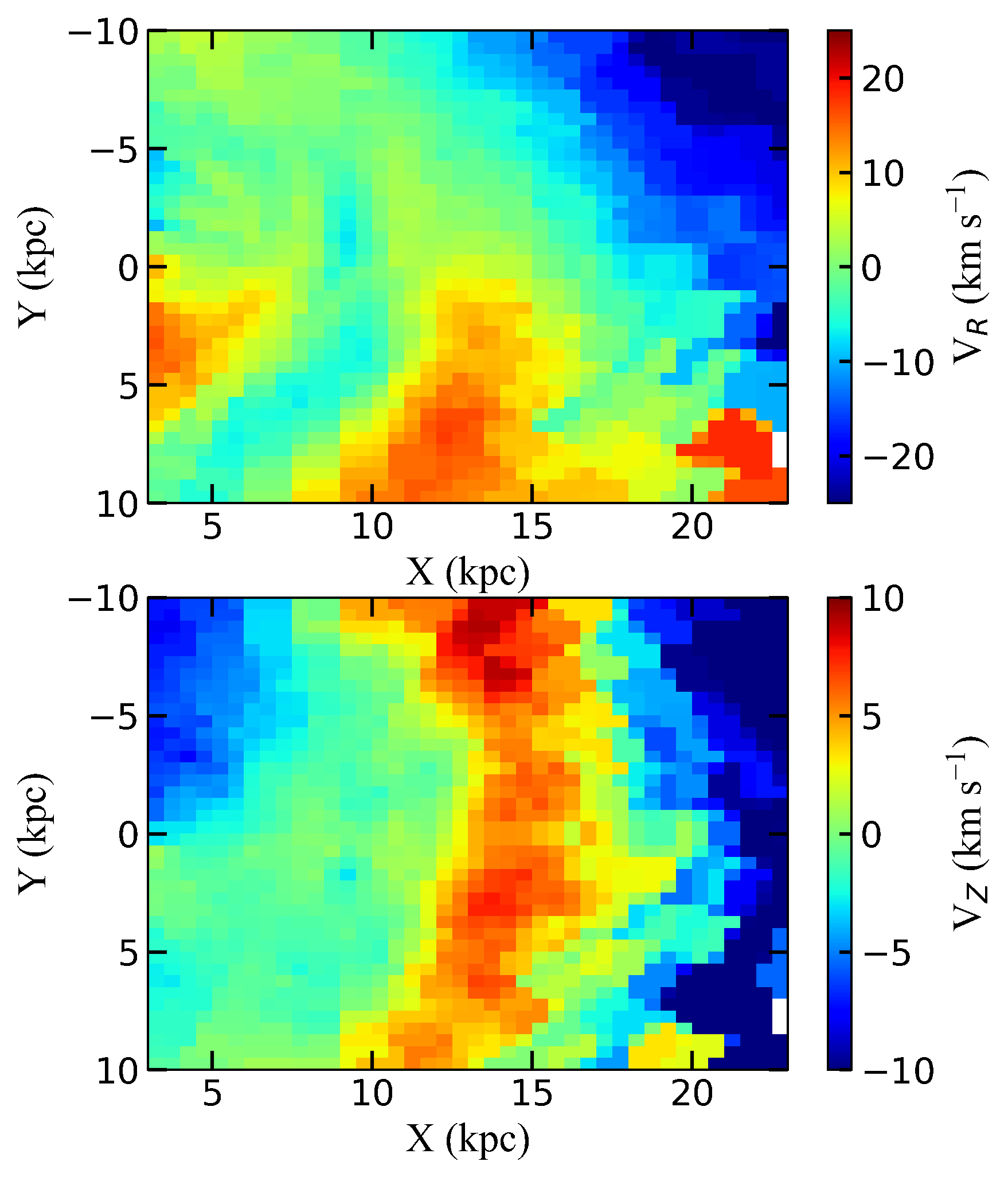}
  \caption{Zoomed version of the radial (upper panel) and vertical 
  (bottom panel) velocity maps shown in  Fig.\ref{lucy2}.}
  \label{zoomvzvR}
\end{figure}
\begin{figure*}
  \centering
  \includegraphics[width=0.8\textwidth]{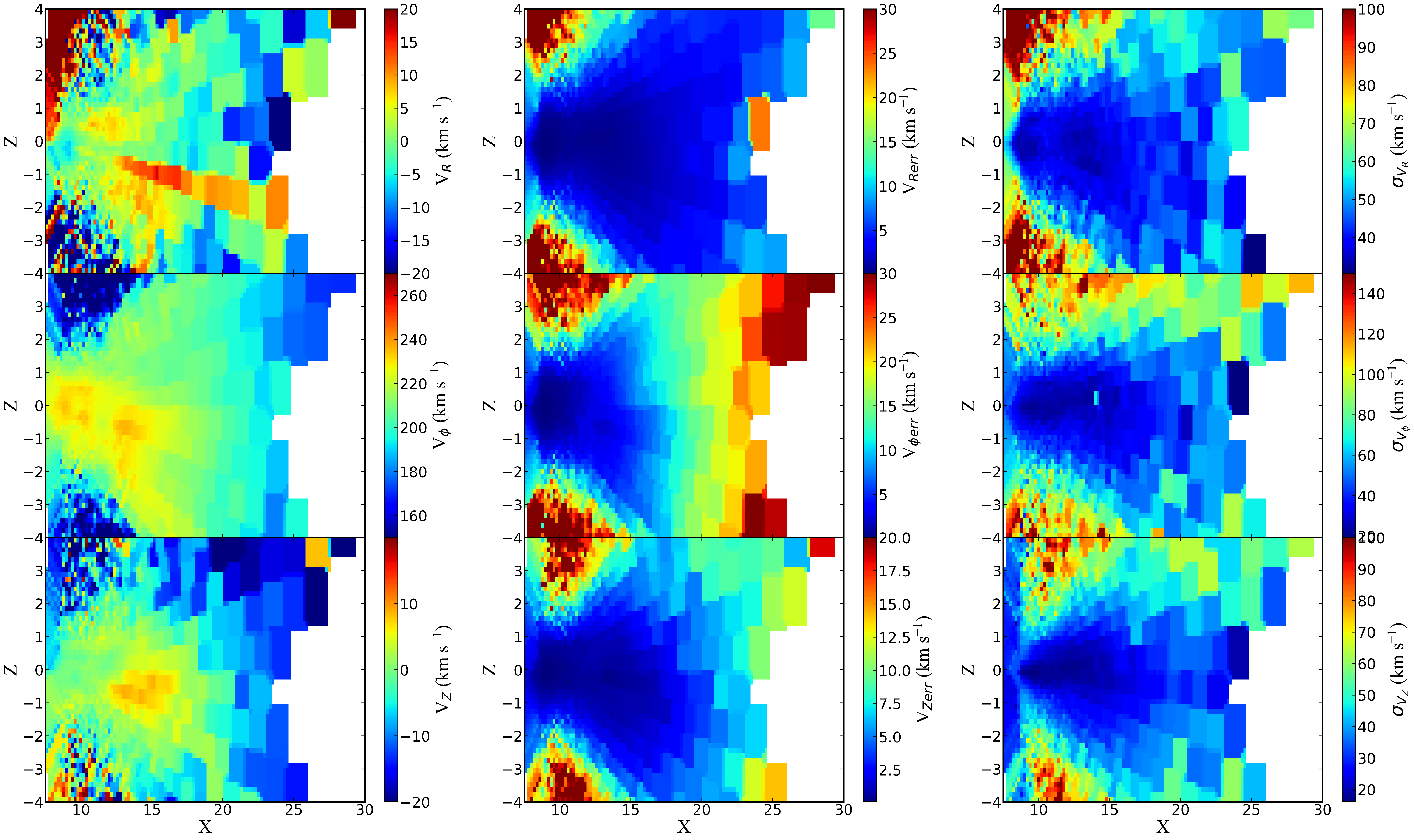}
   \caption{ As  Fig.\ref{lucy2} but the projection is   now on the $(X,Z)$ plane  with the constraints: 
   $160^\circ <\ell <200^\circ $, $\frac{\pi}{\Delta \pi}>1$.
} 
  \label{lucy3}
\end{figure*}

\begin{figure*}
  \centering
  \includegraphics[width=0.8\textwidth]{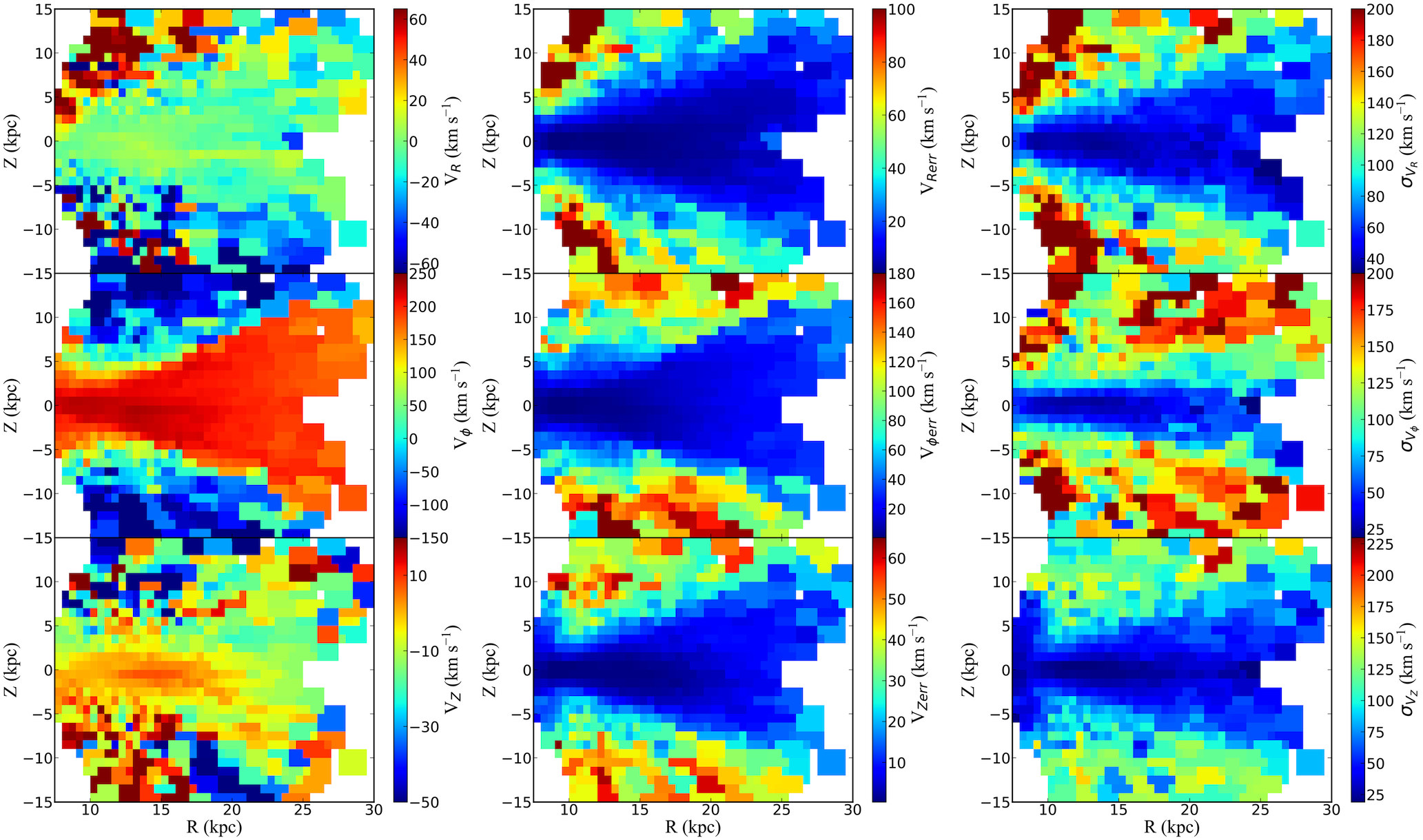}
   \caption{As Fig. \ref{lucy3}, but the projection is on the $(R,Z)$ plane.}
  \label{R-Z-3Dvelocity180}
\end{figure*}

The 
LIM 
gives estimations of the three velocity components, their errors and rms values 
in the $N_{cells}$ cells in which we have divided the galactic region. 
Fig.\ref{lucy2} shows the maps representing the projection 
  onto the Cartesian $(X,Y)$ 
  plane of the  
  three velocity components $V_R, V_\phi$ and $V_Z$ 
  (from top to bottom, left panels), their errors (middle panels) and their rms values (right panels). Note that 
  the deconvolution has included all stars with $\Delta \pi < \pi$ and with  $|b| < 10^\circ$. 

 The Galactic region within  such  constraints  was divided into 36 line-of-sight, each of them with $\Delta \ell = 10^\circ$. 
 Then, in each of these cells we have applied the deconvolution technique  discussed above.
Hereafter only the cells where the number of stars is $N_{stars} \ge 6$  are
plotted. The error 
$\Delta \overline{r}$ in the 
estimation  of the 
distance $\overline r$
is
\be
\label{err1} 
\Delta \overline{r} = \sigma_{\overline{r}} \;.
\ee
where 
$\sigma_{\overline{r}}$
is the dispersion of $\overline{r}$
obtained by the LIM. Note that this 
is a systematic error and  it cannot be reduced by  increasing 
 $N_{stars}$  in each cell.

Errors on the velocity components simply are
\begin{eqnarray}
&&
\Delta V_r =  \frac{\sigma(V_r) }{\sqrt{N_{stars}}} 
\\ \nonumber 
&& 
\Delta \mu_\ell =  \frac{\sigma(\mu_\ell) }{\sqrt{N_{stars}}} 
\\ \nonumber 
&& 
\Delta \mu_b =  \frac{\sigma(\mu_b) }{\sqrt{N_{stars}}}  \;,
\end{eqnarray}
i.e., they
can be reduced by increasing $N_{stars}$.
 
 As discussed in \cite{Lopez-Corredoira_Sylos-Labini_2019} 
  we neglect the covariance
terms in the errors on $\mu_\ell$, $\mu_b$ and $\overline{r}$, 
 that is, we assume these
errors are independent from each other.
Note that 
rms values plotted in the right panels of Fig.\ref{lucy2}  were corrected
(subtracted quadratically) from the measurement errors
of $V_r$, $\mu_\ell$ and $\mu_b$. 
Uncertainties on $V_R$ and $V_\phi$ are smaller
towards the anti-center because the separation of both components
is independent of the distance. Moreover, $V_R$ only depends
on $V_r$, so it is insensitive on the distance errors
which instead affect both 
the determinations of $V_\phi$ and $V_Z$.

We have
tested the impact of the zero-point correction 
given
by \cite{Lindegren_etal_2021}, using the publicly available Python package\footnote{\url{https://gitlab.com/icc-ub/public/gaiadr3_zeropoint}}, which calculates the zero-point as a function of ecliptic latitude, magnitude, and colour. 
By comparing
Fig.\ref{lucy2zp}, that includes the small parallax zero-point bias, with Fig.\ref{lucy2} we can conclude that, in line with similar tests  
presented in \cite{Lopez-Corredoira_Sylos-Labini_2019}, 
such correction  has  very minor effects on the derived maps
 in the region of the anti-center direction, where the detected 
measurements errors are the smallest ones, 
while significant differences are found 
where the errors are larger. However, the region towards the anti-center is the one 
relevant for the analysis of the velocity profiles that we present below.
Indeed, 
one may note  
in  middle panels of Figs. \ref{lucy2}-\ref{lucy2zp} that 
the error distribution 
shows a "horn" like shape, delimiting the region where errors are the
smallest ones: 
the velocity dispersion displays  its lower value along $X$ for small (in absolute value) $Y$,
i.e., the direction of the anti-center.

The azimuthal velocity map (see 2nd row, left panel of Fig. \ref{lucy2})  
displays patterns such that $V_\phi$ decreases with distance in the range [8,25] kpc, while it increases
both for $Y>0$ and $Y<0$.
The vertical 
motions map (see the bottom left panel of Fig.\ref{lucy2}) presents
an  ``arc shape" 
 similar to  the Gaia DR2 
map shown in  \cite{Lopez-Corredoira_Sylos-Labini_2019}:
this is 
now much better resolved,
a fact supporting 
the robustness of the 
LIM.
Similarly, both the behavior of the error and of the dispersion (middle and right panels of Fig. \ref{lucy2}) show the same 
patterns as in the maps obtained from the analysis of DR2.
 Fig.\ref{zoomvzvR} shows a zoomed version of the the radial and vertical velocity  maps presented in  Fig. \ref{lucy2}. 


As seen in Fig. \ref{lucy3}, it shows the edge-on projection, i.e., 
onto the $X,Z$ plane, of the three velocity components
(as in Fig.\ref{lucy2}),
 with the constraint $160^\circ < \ell < 200^\circ$: this 
corresponds to the  galactic anti-center   region.
One may note that there are 
several 
asymmetries with positive and negative velocity gradients of amplitude from 10 to 25 kpc. 
The errors become larger between $Z =$ 2 kpc to 4 kpc and  
between $-$4 kpc to $-$2 kpc and $X \leq$ 15 kpc. 
Note that in Fig. \ref{lucy3} uncertainties 
are  larger  toward the
Galactic pole because  
the binning with constant $\Delta b$  
reduces 
the number of sources in that region.

Fig.~\ref{R-Z-3Dvelocity180} shows the projection on the $(R,Z)$ plane,
where the vertical coordinate has a broader range, i.e. $Z \in [-15,15]$, one
may note asymmetrical kinematics patterns 
such as radial non-null motions, the "horn-like" shape 
in azimuthal velocity and non-zero vertical bulk motions.

\begin{figure}
  \includegraphics[width=0.5\textwidth]{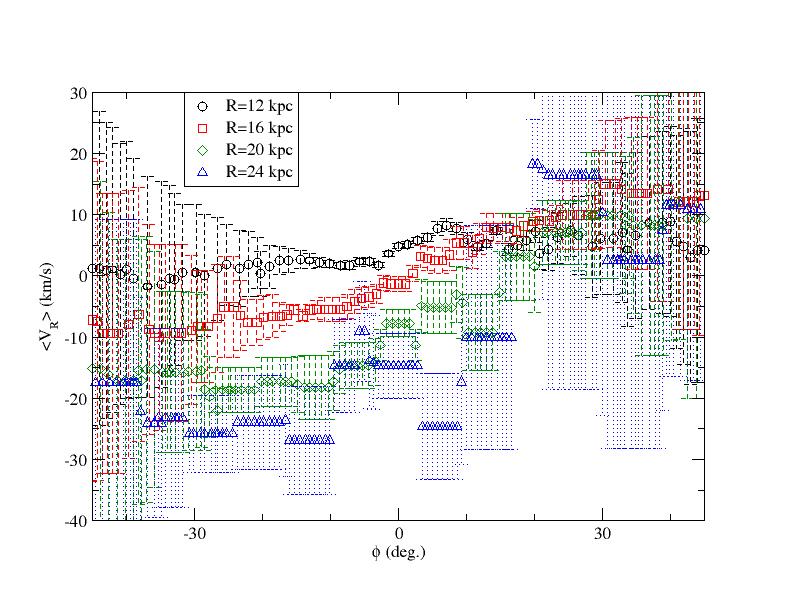}
  \caption{Radial Galactocentric velocity median value 
  as a function of $\phi (\overline{r})$ for different $X(\overline{r})$ 
  (in bins of size $\Delta X=0.5$ kpc) within $|b|<10^\circ $.}
  \label{VRphib10}
\end{figure}

\begin{figure}
  \includegraphics[width=0.5\textwidth]{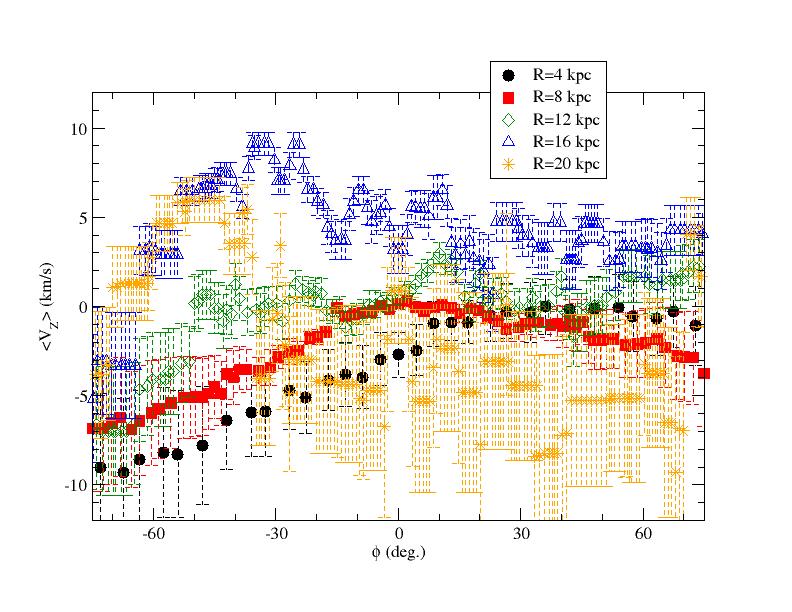}
  \caption{Vertical Galactocentric velocity median value as a function of $\phi (\overline{r})$ for different Galactocentric radii (in bins of size $\Delta R=0.5$ kpc) within $|b|<10^\circ $.
} 
 \label{VZphib10}
\end{figure}

Fig.~\ref{VRphib10} shows the radial velocity profile along the azimuth, 
 in different radial bins and in the range 12$-$24 kpc. 
At least in the range between $-20^\circ$ to $20^\circ$, 
 the larger is the distance and the smaller is
 the radial velocity; beyond this range in azimuth  
 errors are too large to make a reliable estimation of the velocity.
Figure~\ref{VZphib10} shows the vertical velocity profile 
along the azimuth in different radial bins and in the range 4$-$20 kpc. In addition,
$V_Z(\phi)$ increases  from 4 to 16 kpc,
i.e. the larger is the distance the larger is the average vertical velocity.

Finally, the velocity maps and profiles (see below) obtained with the Gaia DR3 are 
consistent with those obtained by \citet{Lopez-Corredoira_Sylos-Labini_2019}
with Gaia DR2, a fact that shows that the LIM is a robust and reliable 
technique. 
This same result can be inferred  by a simple visual comparison of Fig. \ref{lucy2} with Fig. 8 
of \citet{Lopez-Corredoira_Sylos-Labini_2019} 
and of Fig. \ref{lucy3} with their Fig. 9:  
 the 
maps derived from DR3 well agree with those of DR2 even when we consider the
region of DR2 where errors are larger.



\subsection{Asymmetric motions: a qualitative view}

The maps presented in Figs.\ref{lucy2}-\ref{R-Z-3Dvelocity180} 
show the complexity and richness of the velocity field of the Galactic disc. They
again confirm that the Galactic disk is out-of-equilibrium and
it is characterized by asymmetric streaming motions with significant gradients
in all velocity components. 
 These results are in agreement with several others in the literature: indeed 
many 
 velocity substructures, moving groups, bulk motions, radial motions, ridges, snails, arches, etc.,
have been 
revealed in recent works using  star counts and kinematic  
 observations: see, e.g., 
\citet{2012ApJ...750L..41W,Katz_etal_2018,Antoja_etal_2018,2018MNRAS.477.2858W,2020MNRAS.491.2104W,2021antoja,Recio-Blanco_etal_2022,Katz_etal_2022}.

We will present in a forthcoming work a detailed analysis
of the different
dynamical models able to take into account 
the kinematic properties revealed in this analysis. 
We refer the reader for a discussion of a number of theoretical 
possibilities to \cite{2020martin}:
these include
a decomposition of bending and breathing modes, the long bar or bulge, the spiral arms, a tidal interaction with Sagittarius dwarf galaxy, and the analysis of  out-of-equilibrium 
effects.



\subsection{Rotation curve and radial velocity profiles}
\label{rotcurve}

We now discuss the determination of
the radial profiles of the three velocity components  
at different vertical heights. 
We then consider the
derivation of the velocity rotation curve from the Jeans equation, stressing the underlying hypotheses, 
and then we describe in detail our results comparing them, in particular, with those of \cite{Eilers_etal_2019}. 

\subsubsection{Estimation of velocity moments} 

As mentioned above,
the
LIM 
gives estimations of the velocity components and their dispersion, i.e.,
$(V_\alpha^i, \; \sigma_{V_\alpha}^i)$  with $\alpha=R, \phi, Z$ in the $i=1,..,N_{cells}$ cells in which we have divided the galactic region. 
Thus we can estimate the average velocity components as
\be
\label{est1}
\overline{V_\alpha} =  
\frac{\sum_{i}
\frac{V_\alpha^i}{(\sigma_{V_\alpha}^i)^2}  }{\sum_i \frac{1}{(\sigma_{V_\alpha}^i)^2 }} \;.
\ee
Note we compute in bins of size $\Delta R=1$ kpc and at different heights in $Z_{min}, Z_{max}$: for each $R$  
the sums in Eq.\ref{est1} include all the cells such that satisfy such constraints.
The estimation of  variance the of  $\overline{V_X}$ is 
\be
\label{est2}
\sigma_{\overline{V_\alpha}}^2  =  \frac{1}{\sum_i \frac{1}{(\sigma_{V_\alpha}^i)^2 }} 
\ee
with the same constraints as before.

\subsubsection{Results for velocity profile} 

\begin{figure*} 
\includegraphics[width=3.5in]{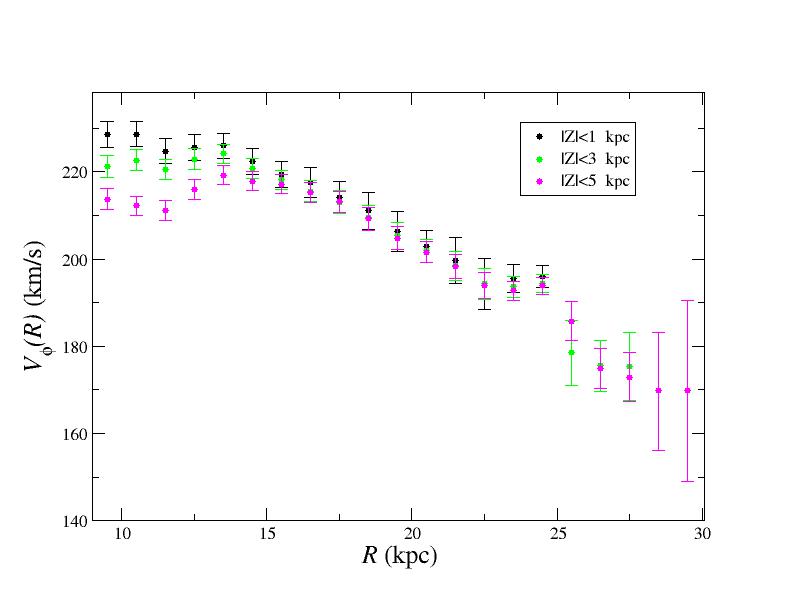}
\includegraphics[width=3.5in]{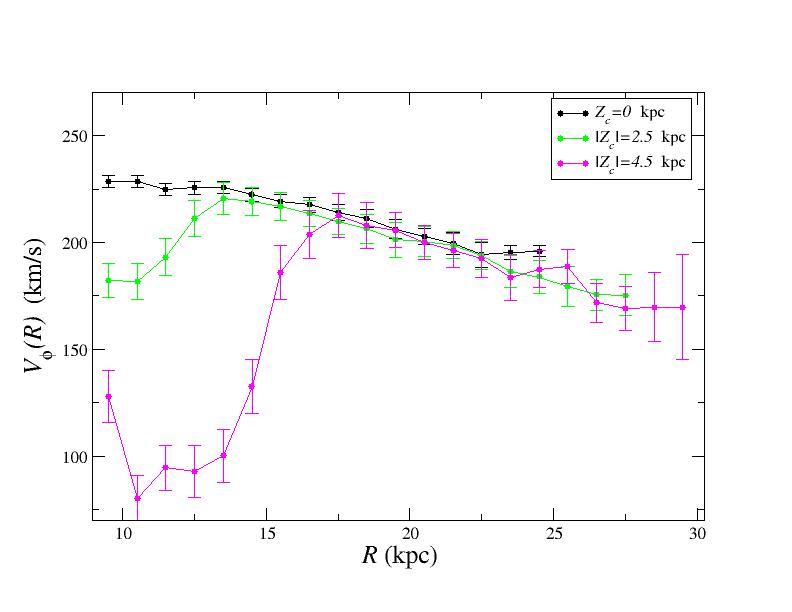}\\
\includegraphics[width=3.5in]{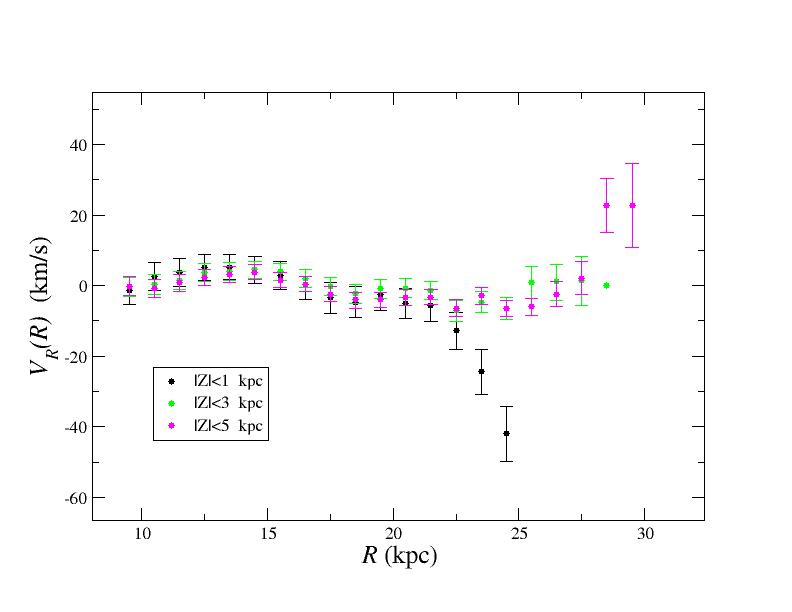}
\includegraphics[width=3.5in]{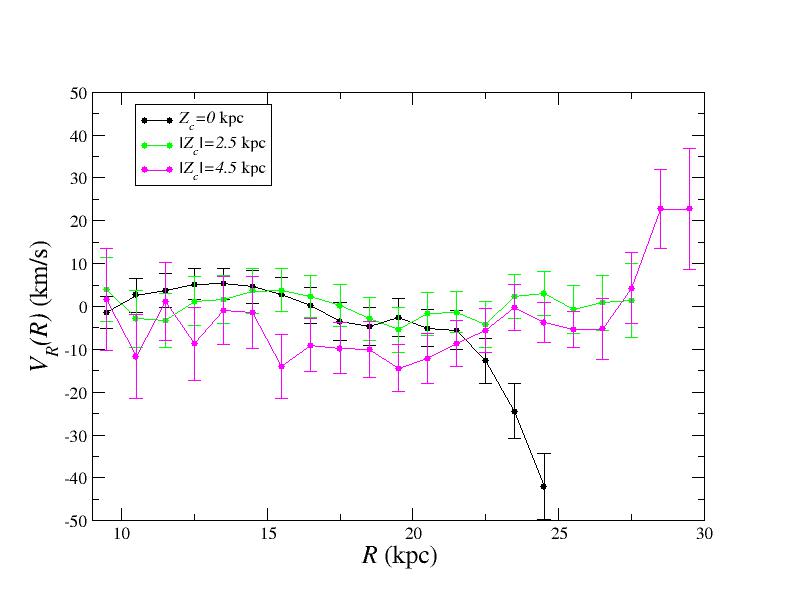}\\
\includegraphics[width=3.5in]{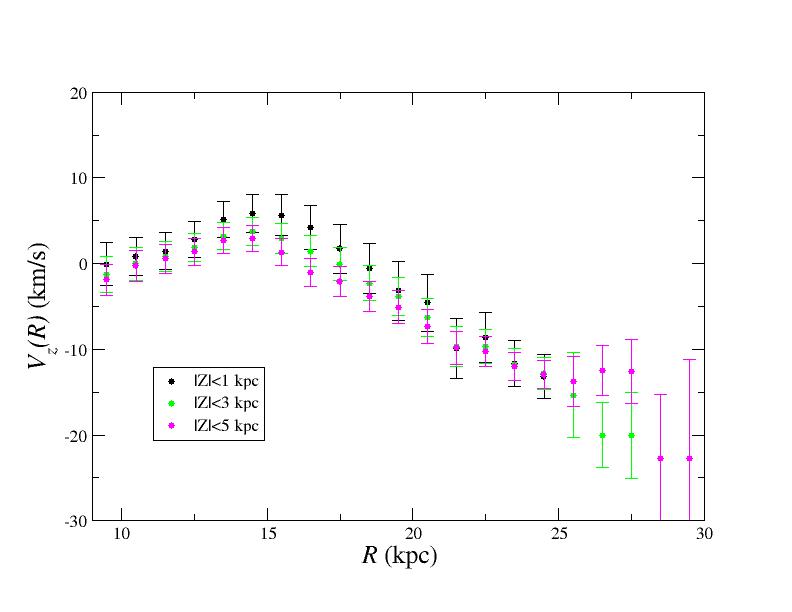}
\includegraphics[width=3.5in]{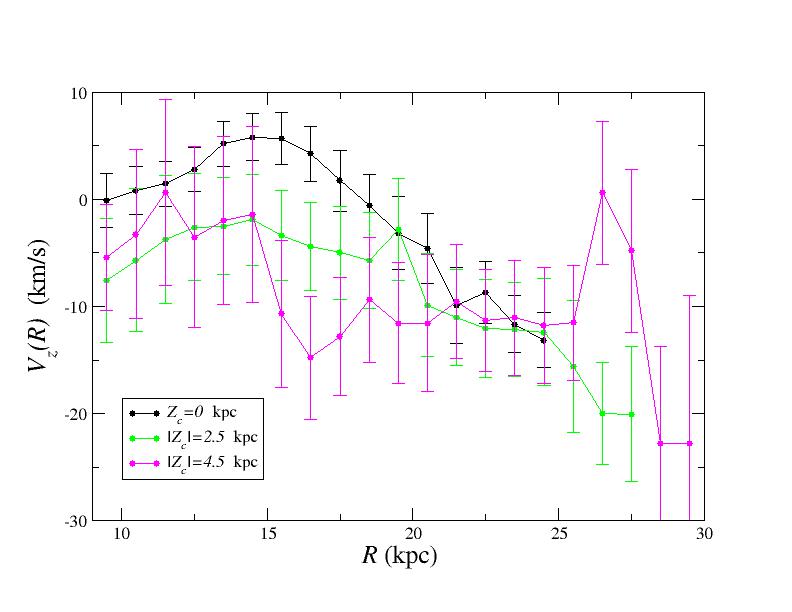}
\caption{From top to bottom: azimuthal, radial and vertical velocity profile
for the  DR3 data-set 
(in the anti-center region  
with the constraint $160^\circ < \ell < 200^\circ$) 
in
different bins 
with vertical height in the range 
$(-Z,Z)$  (left panels) 
and 
in vertical bins 
of size 
$\Delta Z=2$ half centered in $-Z_c$ and half in $Z_c$ 
(right panels).
 }
 \label{V_profile_DR3} 
\end{figure*}

Fig. \ref{V_profile_DR3} shows respectively the transversal, radial and vertical velocity profiles 
along the radial distance and computed in
different vertical slices of size $\Delta Z$ 
(left panels) and 
in bins of size $\Delta Z=1$ centered 
at  
$Z_c$ plus bins of size $\Delta Z=1$ centered  at $-Z_c$
(right panels). 
Hereafter we will refer to these two determinations 
as integral and differential velocity profiles.

The  average azimuthal velocity 
profile
(Fig. \ref{V_profile_DR3} top left panel) shows a clear monotonic decreasing  
trend from  $\sim 220$ km $^{-1}$ at $R \approx 10$ kpc to $\sim 160$ km $^{-1}$ at $R \approx 30$ kpc. The differential determination of the 
azimuthal velocity displays a 
significant trend at small radii, i.e. for $R<15$ kpc $V_\phi$ 
decreases as $|Z_c|$ increases 
(top right panel of Fig. \ref{V_profile_DR3}). 
Instead for $R>15$ kpc, $V_\phi$ 
does not show significant variations
with $Z$. It should be noticed that 
for $|Z|>2$ errors in the determination of the azimuthal velocity are large (up to $\approx 100$ km$^{-1}$ 
--- see rms values in the right panel of Fig. \ref{R-Z-3Dvelocity180})
and thus these results might require 
to be confirmed 
by further data release where 
the error on the parallax will be lowered.

The radial velocity 
profile 
on the Galactic plane (i.e., $Z_c=0$) 
(Fig. \ref{V_profile_DR3} middle  panels)  
decreases in function of radius
for $R<20$ kpc.
Instead, the larger is $Z_c$ and the larger the positive
value of $V_R$, that reaches $\approx 30$ km $^{-1}$
in the outermost regions explored.

Finally, the average vertical velocity  
(Fig. \ref{V_profile_DR3} bottom panels)  
shows a
decreasing trend on the Galactic plane (i.e., $Z_c=0$)
for $R>$
15 kpc 
with a gradient of $20$ km $^{-1}$ until 30 kpc. 

\begin{figure} 
\includegraphics[width=3.5in]{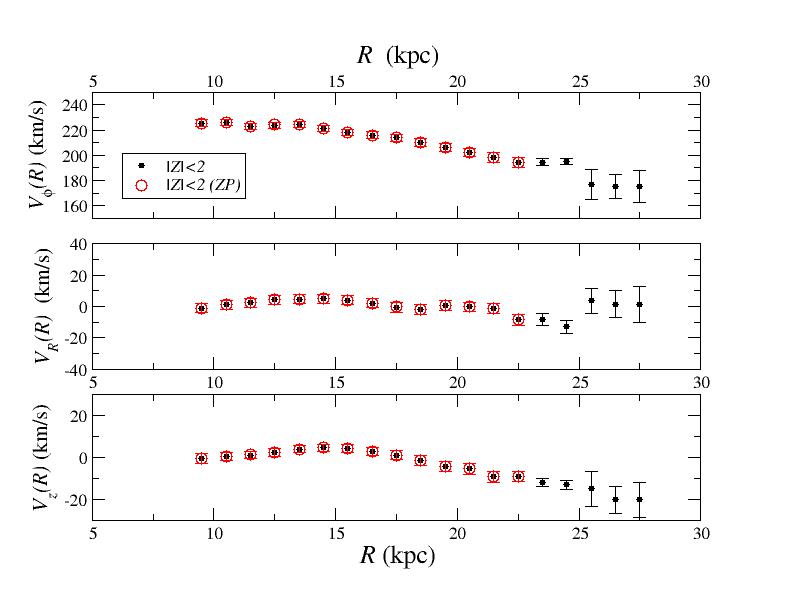}
\caption{Comparison of the profiles with and without the zero-point correction 
computed in the same galactic region.
 }
 \label{V_profile_DR3+ZP} 
\end{figure}

The effect of the zero-point correction on the velocity profiles is shown in Fig.\ref{V_profile_DR3+ZP}. As noticed above the zero-point correction 
affects the behavior at large radii by introducing larger errors and 
for this reason such correction reduces the range of radii where we
can reconstruct kinematic properties. We find that the velocity profiles, when computed 
in the same galactic region, nicely overlap.

\subsubsection{The Jeans Equation}

A basic assumption often used to interpret
Galactic dynamics is that the disk is 
in equilibrium or
that the gravitational potential is stationary: 
 this hypothesis 
is definitely challenged by 
the rich complexity 
of velocity sub-structures reveled by the 
3D kinematic maps provided by the Gaia mission.
Theoretically it is not evident how to take into account 
such streaming motions in all velocity components to construct 
a self-consistent description of the galaxy. In what follows 
we will use the time-independent Jeans equation 
in an axisymmetric gravitational potential
to compute the rotation curve. 
 \cite{Chrobakova_etal_2020}  have shown that,
 as long the amplitude of the radial velocity component is small
 compared to that of the azimuthal one, the Jeans equation
 provides a reasonable approximation to the system. 
 From our analysis we may conclude that up to 30 kpc on the Galactic plane 
 perturbations in the radial velocity should be small and thus we may use the 
Jeans equation
to compare 
observations with theoretical models. 

Assuming an axisymmetric gravitational potential of the Milky Way,
 we use the the Jeans equation (in cylindrical coordinates $R,Z,\phi$) \citep{Binney_Tremaine_2008}, to link the
moments of the velocity distribution and the density of a collective of stars to the gravitational potential, i.e.,
\bea
\label{jeans1}
&&
\frac{ \partial \nu \langle V_R\rangle }{\partial t} 
+
\frac{ \partial \nu \langle V_R^2\rangle }{\partial R} 
+
\frac{ \partial \nu \langle V_R V_Z \rangle }{\partial z}
+ 
\\ \nonumber &&
\nu \left( \frac{ \langle V_R^2\rangle - \langle V_\phi^2\rangle  }{R} 
+  \frac{ \partial \Phi }{\partial R} \right)
=0
\eea
where $\nu$ denotes the density distribution. 

The circular velocity curve in an axisymmetric gravitational potential $\Phi$  of a disk
galaxy is defined as
\be 
\label{vc} 
V^2_c(R) = R  \frac{ \partial \Phi }{\partial R} |_{z \approx 0} \;.
\ee
By assuming a steady state, the time dependent term in Eq.\ref{jeans1} is set to zero and we get 
\be
\label{jeans2}
V_c^2  = \langle V_\phi^2\rangle - \langle V_R^2\rangle 
\left(
1 +
\frac{ \partial \ln \nu  }{\partial \ln R} 
+
\frac{ \partial \ln \langle V_R^2\rangle  }{\partial \ln R} 
\right) 
+ \frac{ \partial \nu \langle V_R V_z \rangle }{\partial z}
\;.
\ee
We assume that the volume density can be written as
\be
\nu(R,z) =  \rho_0 \exp\left( - \frac{R}{h_R} \right) 
\exp\left( - \frac{|z|}{h_z} \right) \;, 
\ee
where $h_R$ is the scale length of the disk and $h_z$ is the scale height
(we use the same values of \cite{Chrobakova_etal_2020}).
By defining for the three velocity components 
\be
\langle V_X^2\rangle  = \langle V_X\rangle^2 + \sigma^2_{\langle V_X\rangle} 
\ee
we find that Eq.\ref{jeans2} can be written as 
 \bea
 \label{vcfinal} 
 &&
 V_c^2 =  \langle V_\phi \rangle^2 +  \sigma^2_{ \langle V_\phi \rangle} + 
 \left( \langle V_R \rangle^2 +  \sigma^2_{ \langle V_R \rangle} \right) \frac{R-h_R}{h_R}
 \\ \nonumber &&
- 2 R \langle V_R \rangle \frac{ \partial  \langle V_R \rangle  }{\partial  R} 
-R \frac{ \partial  \sigma^2_{ \langle V_R \rangle}   }{\partial  R} + 
 \\ \nonumber &&
\frac{R}{h_z}\frac{z}{|z|}  \langle V_R V_z \rangle 
- R\frac{ \partial   \langle V_R V_z \rangle    }{\partial  z} \;.
\eea 

\subsubsection{Rotation curve} 

Following \cite{Eilers_etal_2019} we have neglected the terms with $V_Z$ in Eq. \ref{vcfinal}, 
because the cross-term $\langle V_R V_Z \rangle$  and its vertical gradient is $\approx$ 2-3 orders of magnitude smaller
compared to the remaining terms hence their effects is negligible.

\begin{figure*} 
\includegraphics[width=3.5in]{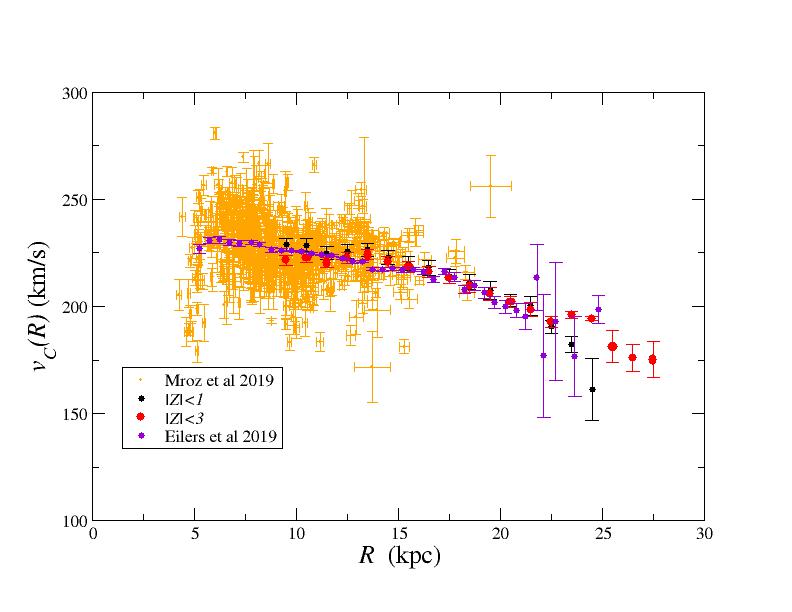}
\includegraphics[width=3.5in]{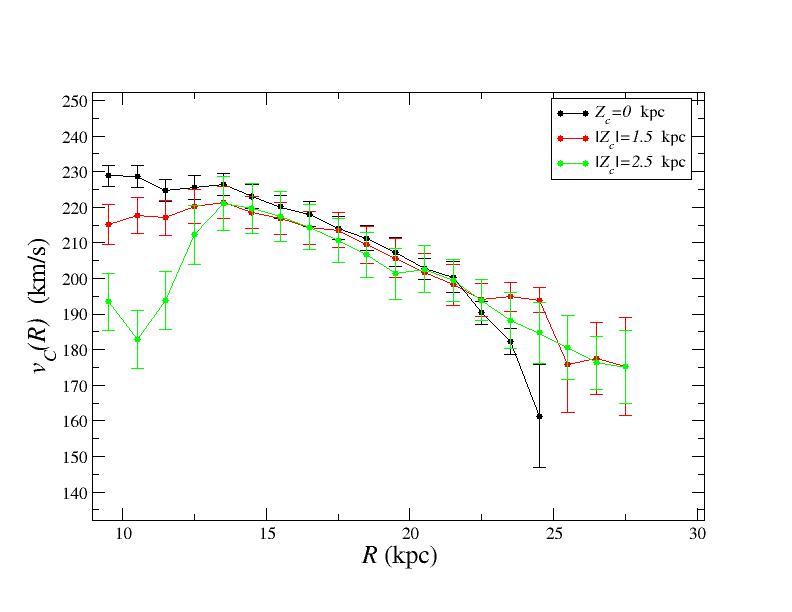}
\caption{
As Fig. \ref{V_profile_DR3} but for 
rotation curve computed in the anti-center region 
with the constraint $160^\circ < \ell < 200^\circ$; limits on |Z| are reported in the labels.
In the left panel data points by \cite{Eilers_etal_2019} 
and by \cite{2019ApJ...870L..10M}  are reported for comparison. }
 \label{VC_profile_DR3} 
\end{figure*}

The resulting 
integral and differential  rotation curve  
is reported in Fig. \ref{VC_profile_DR3} 
the values for $|Z|<$2 kpc are reported in Tab.\ref{tabrot}). 
Note that we limited the analysis to the Galactic plane, 
i.e. $|Z| < 3 $ kpc, i.e., in the region where
 the disc is highly dominant over the stellar halo contribution.

\begin{table} 
\begin{center}
\begin{tabular}{ c c c }
\hline 
\hline 
 $R$ (kpc)  & $v_c$ (km s$^{-1}$)  &  $\sigma_{v_c}$ (km s$^{-1}$)   \\  
\hline 
   9.5  &     221.3&       2.5\\    
   10.5 &     222.6&       2.4\\    
   11.5 &     220.5&       2.4\\    
   12.5 &     222.9&       2.4\\    
   13.5 &     224.1&       2.2\\    
   14.5 &     220.7&       2.3\\    
   15.5 &     218.1&       2.3\\   
   16.5 &     215.5&       2.4\\    
   17.5 &     213.0&       2.6\\    
   18.5 &     209.4&       2.9\\    
   19.5 &     205.4&       3.1\\    
   20.5 &     201.8&       2.7\\    
   21.5 &     198.4&       3.3\\    
   22.5 &     194.3&       3.5\\    
   23.5 &     193.7&       2.4\\    
   24.5 &     194.4&       2.1\\    
   25.5 &     178.5&       7.4\\    
   26.5 &     175.5&       5.8\\    
   27.5 &     175.3&       7.7\\     
\hline
\end{tabular}
\end{center}
\caption{Measurements of the circular velocity of the Milky Way 
for the Gaia DR3 sample in the anti-center region, i.e. 
with the constraints $160^\circ < \ell < 200^\circ$ and $|Z|<$3 kpc. 
Columns show the Galactocentric radius, the circular velocity, and its
 error bar.}
\label{tabrot} 
\end{table}

By comparing $V_C$ 
with the azimuthal velocity profile 
(upper panel of Fig. \ref {V_profile_DR3})
we notice that the additional terms 
from the Jeans equation only contributes as small perturbations.   
In particular, the main features that we have observed for 
the azimuthal velocity profile are present also for the rotation curve.
Namely, the decrease in amplitude passing from 10 kpc to 30 kpc 
and, when we consider the differential 
measurement, the trend of the decreasing amplitude at small radii 
with the increase (in absolute value) of the vertical height.
As mentioned above (see Fig.\ref{V_profile_DR3+ZP}),
we have tested that the systematic contribution 
due to the zero-point correction does not change significantly the behaviors 
observed.

That the rotation curve of the Milky Way was  decreasing for $R>12$ kpc 
was also found by several authors with different tracers and modelling (see, e.g.,  \citet{Dias+Lepine_2005,2008ApJ...684.1143X,Bovy_etal_2012,Kafle_etal_2012,Reid_etal_2014,2014MNRAS.445.3788G,Lopez-Corredoira_2014,GalazutdinoV_etal_2015,Katz_etal_2018,Lopez-Corredoira_Sylos-Labini_2019,2019ApJ...870L..10M,Eilers_etal_2019,2021A&A...654A..25J,2022MNRAS.510.2242W,2022MNRAS.516..731B}).
In particular, 
\cite{Eilers_etal_2019}
considered a sample
with the 6D   phase-space coordinates of
23,000 luminous red giant stars, with precise parallaxes determined 
by combining spectral data from APOGEE DR14 with photometric information from WISE, 2MASS, and Gaia DR2.
They measured that the circular velocity curve shows a gentle
but significant decline with increasing radius and can be well
approximated by a linear function up to 25 kpc:
\be
\label{rc}
V_C(R)= V(R_\odot)  + \beta (R - R_\odot) 
\ee
where $R_\odot$ is the distance of the Sun from the 
Galactic center, $V(R_\odot)=229 \pm 0.2  \; \mbox{km s}^{-1}$  and
the slope was found to be 
$\beta=-(1.7 \pm 0.1)  \;  \mbox{km s}^{-1} \mbox{kpc}^{-1}$. 
In Fig. \ref {V_profile_DR3} we report 
the rotation curve determined by \citet{Eilers_etal_2019} which nicely agrees with our estimation:
in turn their results are in good agreement with previous determinations by 
\cite{Kafle_etal_2012,Lopez-Corredoira_2014,Huang_etal_2016} 
although these three determinations have larger error bars (see  Fig.3
of \cite{Eilers_etal_2019}).

The result of \cite{Eilers_etal_2019} is in reasonably good agreement with 
another recent analysis of the Milky Way’s circular
velocity curve by \cite{2019ApJ...870L..10M} (also reported 
in Fig.\ref {V_profile_DR3} for comparison). This was based on  
a sample of 773 Classical Cepheids with precise distances based on mid-infrared period–luminosity relations coupled with proper motions and radial velocities from Gaia. 
They found,   
in the range of radii between 5 kpc and 20 kpc, a somewhat
larger slope of  $\beta=-(1.4 \pm 0.1) \; \mbox{km s}^{-1} \mbox{kpc}^{-1}$. Note from Fig.\ref {V_profile_DR3} that, however,  
the number of Cepheids for $R>15$ kpc drops significantly so that 
the slope $\beta$ was measured on a very limited range of radii.  

By extending the rotation curve to 27.5 kpc we find that 
$\beta=-(2.3 \pm 0.2)  \; \mbox{km s}^{-1} \mbox{kpc}^{-1}$, which is
smaller than the other two determinations mentioned above. 
However, in the range of radii
where they overlap, i.e. for $R<20$ kpc for the 
determination by \cite{2019ApJ...870L..10M} and for $R<25$ kpc for the one by 
\cite{Eilers_etal_2019}, the three measurements are 
in reasonably good agreement with each other. 
In this respect we note that our result is independent from those  of
\cite{2019ApJ...870L..10M} and \cite{Eilers_etal_2019} as we used a different 
data-set and a different method to determine the rotation curve. Namely, 
we have constructed a coarse-grained sample and we have applied to it a statistical method
to measure the average velocity components and their dispersion,
which was tested to provide reliable results,
rather than analyzing the  velocities of individual stars.

In order to estimate  systematic uncertainties on the
circular velocity curve arising from our data sample, we split
the  galactic region into two disjoint smaller portions,
one with $0$ kpc $<Z<$ 2 kpc and the other one with 
-2 kpc $<Z<$ 0 kpc (a similar result 
is obtained by dividing the sample $b>0^\circ$ and $b<0^\circ$). 
We have then computed the rotation curve
in the two disjointed regions and we have 
estimated the systematic uncertainties on the circular velocity by
the difference between the resulting fit parameters from the two
disjoint data sets. 
We find that 
the systematic error on the slope of the rotation curve $\beta$ (see Eq.\ref{rc})
is
$\approx 0.4$ km s$^{-1}$ kpc$^{-1}$ which corresponds 
to a uncertainty of the order of 20\%.
An additional contribution to the systematic 
comes from the fact that,
for calculating the circular velocity curve from Eq.\ref{jeans2},
we neglected the cross
terms $\langle V_z V_R \rangle$. 
The large noise affecting this terms does not allow a precise determination 
of its effect, but in agreement with \cite{Eilers_etal_2019} we estimate 
it of the order of a few percent.

\subsubsection{Discussion}

The two determinations of the Galaxy rotation curve by \cite{2019ApJ...870L..10M,Eilers_etal_2019}
and the one presented in this work (that up to 20 kpc coincides  
with the one by \cite{Lopez-Corredoira_Sylos-Labini_2019})

are different  
from others  reported in the literature (see, e.g. \cite{2014ApJ...785...63B,2020Galax...8...37S})
where the rotation curve of the Milky Way did not present 
a decrease in the range of distances between 15 kpc and 30 kpc. 
In such estimations 
$V_C(R)$ was measured when, except in few cases,
the full three-dimensional velocity information of the tracers
was not available, and it has to be reconstructed from
only the measured line-of-sight velocity and positional
information of various tracer objects in the Galaxy.
On the other hand,  results 
by \cite{2019ApJ...870L..10M},  \cite{Eilers_etal_2019} and by us in this work 
share the key fact that they are based on different measurements 
that have an unprecedented precision and accuracy in the determination of the distances 
and that allow to have information on the 6D phase-space. The knowledge of the 
phase-space distribution allows to control for possible systematic effects of different kind as 
for instance that 
induced by coherent radial and vertical motions. 
Given the agreement between these determinations we conclude that our results are reliable and that possible systematic effects in the determination of the rotation curve should be marginal: of course, when the next data release of the Gaia mission will be published it will be possible to further test the rotation curve for $R>20$ kpc. This situation actually shows that 
the  better accuracy of the stars distances
together with the great amount of new information provided by 
the Gaia mission is the key improvement to our knowledge of the Galaxy,
its kinematics and dynamics. 

Concerning the estimation of mass of the Milky Way we note that 
\cite{Eilers_etal_2019}, by considering the 
standard Navarro Frenk and White halo model, found 
a virial mass 
of $M = (7.25 \pm 0.25) \times 10^{11} M_\odot$ 
which is significantly lower than what
several previous studies suggest. 
The declining trend 
of the  rotation curve continues when we extend 
the range of radii  form 25 kpc to 27.5 kpc, 
implies that the best fit with a Navarro, Frenk and White halo model \citep{1997ApJ...490..493N} 
should correspond to an even smaller Milky Way virial mass. 
A detailed discussion of the dynamical 
implication of these results will be presented in a forthcoming publication.

Another noticeable feature of the circular velocity
that we have detected, namely the trend of the decreasing amplitude for $R<15$ kpc 
with the increase (in absolute value) of the vertical height has, 
to our knowledge, not been noticed
before. 
A similar study was presented in \cite{Chrobakova_etal_2020} with Gaia DR2, 
reaching distance up to $R\approx20$ kpc and heights only up to $|Z|<2$ kpc, where these
trends are less noticeable. 
Moreover, the binning of data in \cite{Chrobakova_etal_2020} was  more finer, meaning that the derivatives in Eq. (\ref{vcfinal}) are more influenced by fluctuations, 
making them less reliable. Thus the rotation curves of \cite{Chrobakova_etal_2020} are less robust than our current analysis 
and the trends we see now were not noticed then. 
As for the case of the determination of $V_\phi(R,Z)$ 
we emphasize that 
for $|Z|>2$ kpc errors 
in the determination of the azimuthal velocity 
are large (up to $\approx 100$ km$^{-1}$ 
--- see rms values in the right panel of  Fig. \ref{R-Z-3Dvelocity180}).

\section{Conclusion}
 \label{sect5}

We have used the Lucy's inversion model (LIM) to analyze the Gaia DR3 data-set \citep{Gaia_DR3}. 
The LIM  can solve the deconvolution of large Gaussian errors that affect the measurements of stellar distance
and it was previously applied to the DR2 data-set \citep{Lopez-Corredoira_Sylos-Labini_2019}: in this way it 
was possible to derive the 
kinematic maps of the Galaxy covering a  region where
the relative error in distance was larger than 20 \%
thus 
extending the range of distances for the kinematic analyses with respect to those
 presented by  \cite{Katz_etal_2018}. 
The new analysis discussed in the present work allow us to 
explore a range of Galactocentric distances up to $\approx$ 30 kpc,
while the range of distances
covered by studies using only stars with distance errors  $<20\%$
also  increased, by passing from DR2 to 
 DR3, from 13 kpc 
 \citep{Katz_etal_2018}  
 to 18 kpc \citep{Gaia_DR3_K}. 

The first noticeable results is that the LIM 
 applied to the DR3 data 
are compatible with the results of DR2 \cite{Lopez-Corredoira_Sylos-Labini_2019}. 
This is already an important
result showing that the LIM method converges: that is, by lowering the 
parallax errors and by increasing the number of sources, i.e. 
the measured stars. The results well confirm those obtained
earlier in a data-set with less objects and larger errors.
As the method is designed to work when the distribution
of error is Gaussian, this means that the observational
parallaxes error satisfy such condition. 

In addition, as second key result of our work, we find that the new 
extended maps of the Galactic disk cover the regions of the outer disk that are
farther from the Galactic center, whose stars reach 
$R\approx 30$ kpc \citep{2018martin}. These maps show that
there are large amplitude and coherent streaming motions in all velocity components.
In particular, the radial velocity profile shows an increase toward
the outermost region of the Galaxy, but off plane,
with a detected value of $V_R \approx 30$ km s$^{-1}$. 
The azimuthal velocity also shows a clear decreasing trend 
in function of the radius. In addition, we have found a marked 
change of $V_\phi(R)$ at small radii, i.e. $R<15$ kpc,
when we consider its determination at different heights. 
 In summary, the new extended maps confirm that the Galaxy kinematics is 
 characterized by significant coherent streaming motions in 
 all velocity components as found, at smaller radii,  by, e.g., 
 \cite{Katz_etal_2018,Antoja_etal_2018,Lopez-Corredoira_Sylos-Labini_2019,Khoperskov_etal_2021}.

By computing the 
rotation curve through the Jeans equation, assuming that the Galaxy
is in a steady state and that the galactic potential is axisymmetric, 
we found that $V_c(R)$ ($V_\phi(R)$) shows a significant decline from 
$\approx 15$ kpc  to 30 kpc of more than 50 km s $^{-1}$. 
This result is in reasonable agreement with the recent findings by \cite{2019ApJ...870L..10M} up to 20 kpc and by
\cite{Eilers_etal_2019} up to 25 kpc 
and extends them to 27.5
kpc but with smaller errors.
(Of course, the behavior of the rotation curve in this paper 
agrees with that found by \cite{Lopez-Corredoira_Sylos-Labini_2019} 
by applying the LIM on the Gaia DR2 sources).
These three results used different samples  
that have an unprecedented precision and accuracy 
in the determination of the distances 
and that allow to have information on the 6D phase space.
Another interesting result that we found is that the rotation curve, as well the azimuthal velocity, presents a marked dependence on the height for $R<15$ kpc 
whereas, in both cases, at larger $R$ the dependence on $Z$ is negligible.
In a forthcoming work we will interpret these behaviors 
at the light of different galactic models providing an estimation
of the Galaxy's mass.

\section*{Acknowledgments}
We would like to thank 
Roberto Capuzzo-Dolcetta very much for useful comments and insightful discussion.  HFW acknowledges the support from the project  "Complexity in self-gravitating systems", of the Enrico Fermi Research Center (Rome, Italy) and the science research grants from the China Manned Space Project with NO. CMS-CSST-2021-B03, CMS-CSST-2021-A08. ZC was supported by the VEGA – the Slovak Grant Agency for Science, grant No. 1/0761/21 and by the Erasmus+ programe of the European Union under grant No. 2020-1-CZ01-KA203-078200.
This work has made use of data from the European Space Agency (ESA) mission
{\it Gaia} (\url{https://www.cosmos.esa.int/gaia}), processed by the {\it Gaia}
Data Processing and Analysis Consortium (DPAC,
\url{https://www.cosmos.esa.int/web/gaia/dpac/consortium}). Funding for the DPAC
has been provided by national institutions, in particular the institutions
participating in the {\it Gaia} Multilateral Agreement.

\bibliography{HFbibliography.bib}

\begin{thebibliography}{}
\expandafter\ifx\csname natexlab\endcsname\relax\def\natexlab#1{#1}\fi
\providecommand{\url}[1]{\href{#1}{#1}}
\providecommand{\dodoi}[1]{doi:~\href{http://doi.org/#1}{\nolinkurl{#1}}}
\providecommand{\doeprint}[1]{\href{http://ascl.net/#1}{\nolinkurl{http://ascl.net/#1}}}
\providecommand{\doarXiv}[1]{\href{https://arxiv.org/abs/#1}{\nolinkurl{https://arxiv.org/abs/#1}}}

\bibitem[{{Antoja} {et~al.}(2017){Antoja}, {de Bruijne}, {Figueras}, {Mor},
  {Prusti}, \& {Roca-F{\`a}brega}}]{2017antoja}
{Antoja}, T., {de Bruijne}, J., {Figueras}, F., {et~al.} 2017, \aap, 602, L13,
  \dodoi{10.1051/0004-6361/201731060}

\bibitem[{{Antoja} {et~al.}(2016){Antoja}, {Roca-F{\`a}brega}, {de Bruijne}, \&
  {Prusti}}]{Antoja2016}
{Antoja}, T., {Roca-F{\`a}brega}, S., {de Bruijne}, J., \& {Prusti}, T. 2016,
  \aap, 589, A13, \dodoi{10.1051/0004-6361/201628200}

\bibitem[{{Antoja} {et~al.}(2018){Antoja}, {Helmi}, {Romero-G{\'o}mez}, {Katz},
  {Babusiaux}, {Drimmel}, {Evans}, {Figueras}, {Poggio}, {Reyl{\'e}}, {Robin},
  {Seabroke}, \& {Soubiran}}]{Antoja_etal_2018}
{Antoja}, T., {Helmi}, A., {Romero-G{\'o}mez}, M., {et~al.} 2018, Nature, 561,
  360, \dodoi{10.1038/s41586-018-0510-7}

\bibitem[{{Antoja} {et~al.}(2021){Antoja}, {McMillan}, {Kordopatis}, {Ramos},
  {Helmi}, {Balbinot}, {Cantat-Gaudin}, {Chemin}, {Figueras}, {Jordi},
  {Khanna}, {Romero-G{\'o}mez}, {Seabroke}, {Brown}, {Vallenari}, {Prusti}, {de
  Bruijne}, {Babusiaux}, {Biermann}, {Creevey}, {Evans}, {Eyer}, {Hutton},
  {Jansen}, {Klioner}, {Lammers}, {Lindegren}, {Luri}, {Mignard}, {Panem},
  {Pourbaix}, {Randich}, {Sartoretti}, {Soubiran}, {Walton}, {Arenou},
  {Bailer-Jones}, {Bastian}, {Cropper}, {Drimmel}, {Katz}, {Lattanzi}, {van
  Leeuwen}, {Bakker}, {Casta{\~n}eda}, {De Angeli}, {Ducourant}, {Fabricius},
  {Fouesneau}, {Fr{\'e}mat}, {Guerra}, {Guerrier}, {Guiraud}, {Jean-Antoine
  Piccolo}, {Masana}, {Messineo}, {Mowlavi}, {Nicolas}, {Nienartowicz},
  {Pailler}, {Panuzzo}, {Riclet}, {Roux}, {Sordo}, {Tanga}, {Th{\'e}venin},
  {Gracia-Abril}, {Portell}, {Teyssier}, {Altmann}, {Andrae}, {Bellas-Velidis},
  {Benson}, {Berthier}, {Blomme}, {Brugaletta}, {Burgess}, {Busso}, {Carry},
  {Cellino}, {Cheek}, {Clementini}, {Damerdji}, {Davidson}, {Delchambre},
  {Dell'Oro}, {Fern{\'a}ndez-Hern{\'a}ndez}, {Galluccio}, {Garc{\'\i}a-Lario},
  {Garcia-Reinaldos}, {Gonz{\'a}lez-N{\'u}{\~n}ez}, {Gosset}, {Haigron},
  {Halbwachs}, {Hambly}, {Harrison}, {Hatzidimitriou}, {Heiter},
  {Hern{\'a}ndez}, {Hestroffer}, {Hodgkin}, {Holl}, {Jan{\ss}en}, {Jevardat de
  Fombelle}, {Jordan}, {Krone-Martins}, {Lanzafame}, {L{\"o}ffler}, {Lorca},
  {Manteiga}, {Marchal}, {Marrese}, {Moitinho}, {Mora}, {Muinonen}, {Osborne},
  {Pancino}, {Pauwels}, {Recio-Blanco}, {Richards}, {Riello}, {Rimoldini},
  {Robin}, {Roegiers}, {Rybizki}, {Sarro}, {Siopis}, {Smith}, {Sozzetti},
  {Ulla}, {Utrilla}, {van Leeuwen}, {van Reeven}, {Abbas}, {Abreu Aramburu},
  {Accart}, {Aerts}, {Aguado}, {Ajaj}, {Altavilla}, {{\'A}lvarez}, {{\'A}lvarez
  Cid-Fuentes}, {Alves}, {Anderson}, {Varela}, {Audard}, {Baines}, {Baker},
  {Balaguer-N{\'u}{\~n}ez}, {Balog}, {Barache}, {Barbato}, {Barros}, {Barstow},
  {Bartolom{\'e}}, {Bassilana}, {Bauchet}, {Baudesson-Stella}, {Becciani},
  {Bellazzini}, {Bernet}, {Bertone}, {Bianchi}, {Blanco-Cuaresma}, {Boch},
  {Bombrun}, {Bossini}, {Bouquillon}, {Bragaglia}, {Bramante}, {Breedt},
  {Bressan}, {Brouillet}, {Bucciarelli}, {Burlacu}, {Busonero}, {Butkevich},
  {Buzzi}, {Caffau}, {Cancelliere}, {C{\'a}novas}, {Carballo}, {Carlucci},
  {Carnerero}, {Carrasco}, {Casamiquela}, {Castellani}, {Castro-Ginard},
  {Castro Sampol}, {Chaoul}, {Charlot}, {Chiavassa}, {Cioni}, {Comoretto},
  {Cooper}, {Cornez}, {Cowell}, {Crifo}, {Crosta}, {Crowley}, {Dafonte},
  {Dapergolas}, {David}, {David}, {de Laverny}, {De Luise}, {De March}, {De
  Ridder}, {de Souza}, {de Teodoro}, {de Torres}, {del Peloso}, {del Pozo},
  {Delgado}, {Delgado}, {Delisle}, {Di Matteo}, {Diakite}, {Diener},
  {Distefano}, {Dolding}, {Eappachen}, {Enke}, {Esquej}, {Fabre}, {Fabrizio},
  {Faigler}, {Fedorets}, {Fernique}, {Fienga}, {Fouron}, {Fragkoudi}, {Fraile},
  {Franke}, {Gai}, {Garabato}, {Garcia-Gutierrez}, {Garc{\'\i}a-Torres},
  {Garofalo}, {Gavras}, {Gerlach}, {Geyer}, {Giacobbe}, {Gilmore}, {Girona},
  {Giuffrida}, {Gomez}, {Gonzalez-Santamaria}, {Gonz{\'a}lez-Vidal}, {Granvik},
  {Guti{\'e}rrez-S{\'a}nchez}, {Guy}, {Hauser}, {Haywood}, {Hidalgo}, {Hilger},
  {H{\l}adczuk}, {Hobbs}, {Holland}, {Huckle}, {Jasniewicz}, {Jonker},
  {Juaristi Campillo}, {Julbe}, {Karbevska}, {Kervella}, {Kochoska},
  {Kontizas}, {Korn}, {Kostrzewa-Rutkowska}, {Kruszy{\'n}ska}, {Lambert},
  {Lanza}, {Lasne}, {Le Campion}, {Le Fustec}, {Lebreton}, {Lebzelter},
  {Leccia}, {Leclerc}, {Lecoeur-Taibi}, {Liao}, {Licata}, {Lindstr{\o}m},
  {Lister}, {Livanou}, {Lobel}, {Madrero Pardo}, {Managau}, {Mann}, {Marchant},
  {Marconi}, {Marcos Santos}, {Marinoni}, {Marocco}, {Marshall}, {Martin Polo},
  {Mart{\'\i}n-Fleitas}, {Masip}, {Massari}, {Mastrobuono-Battisti}, {Mazeh},
  {Messina}, {Michalik}, {Millar}, {Mints}, {Molina}, {Molinaro}, {Moln{\'a}r},
  {Montegriffo}, {Mor}, {Morbidelli}, {Morel}, {Morris}, {Mulone}, {Munoz},
  {Muraveva}, {Murphy}, {Musella}, {Noval}, {Ord{\'e}novic}, {Orr{\`u}},
  {Osinde}, {Pagani}, {Pagano}, {Palaversa}, {Palicio}, {Panahi}, {Pawlak},
  {Pe{\~n}alosa Esteller}, {Penttil{\"a}}, {Piersimoni}, {Pineau}, {Plachy},
  {Plum}, {Poggio}, {Poretti}, {Poujoulet}, {Pr{\v{s}}a}, {Pulone}, {Racero},
  {Ragaini}, {Rainer}, {Raiteri}, {Rambaux}, {Ramos-Lerate}, {Re Fiorentin},
  {Regibo}, {Reyl{\'e}}, {Ripepi}, {Riva}, {Rixon}, {Robichon}, {Robin},
  {Roelens}, {Rohrbasser}, {Rowell}, {Royer}, {Rybicki}, {Sadowski},
  {Sagrist{\`a} Sell{\'e}s}, {Sahlmann}, {Salgado}, {Salguero}, {Samaras},
  {Sanchez Gimenez}, {Sanna}, {Santove{\~n}a}, {Sarasso}, {Schultheis},
  {Sciacca}, {Segol}, {Segovia}, {S{\'e}gransan}, {Semeux}, {Siddiqui},
  {Siebert}, {Siltala}, {Slezak}, {Smart}, {Solano}, {Solitro}, {Souami},
  {Souchay}, {Spagna}, {Spoto}, {Steele}, {Steidelm{\"u}ller}, {Stephenson},
  {S{\"u}veges}, {Szabados}, {Szegedi-Elek}, {Taris}, {Tauran}, {Taylor},
  {Teixeira}, {Thuillot}, {Tonello}, {Torra}, {Torra}, {Turon}, {Unger},
  {Vaillant}, {van Dillen}, {Vanel}, {Vecchiato}, {Viala}, {Vicente},
  {Voutsinas}, {Weiler}, {Wevers}, {Wyrzykowski}, {Yoldas}, {Yvard}, {Zhao},
  {Zorec}, {Zucker}, {Zurbach}, \& {Zwitter}}]{Antoja2021}
{Antoja}, T., {McMillan}, P.~J., {Kordopatis}, G., {et~al.} 2021, \aap, 649,
  A8, \dodoi{10.1051/0004-6361/202039714}

\bibitem[{{Bennett} \& {Bovy}(2019)}]{Bennett2019}
{Bennett}, M., \& {Bovy}, J. 2019, \mnras, 482, 1417,
  \dodoi{10.1093/mnras/sty2813}

\bibitem[{{Bhattacharjee} {et~al.}(2014){Bhattacharjee}, {Chaudhury}, \&
  {Kundu}}]{2014ApJ...785...63B}
{Bhattacharjee}, P., {Chaudhury}, S., \& {Kundu}, S. 2014, \apj, 785, 63,
  \dodoi{10.1088/0004-637X/785/1/63}

\bibitem[{Binney \& Tremaine(2008)}]{Binney_Tremaine_2008}
Binney, J., \& Tremaine, S. 2008, Galactic Dynamics (Princeton University
  Press)

\bibitem[{{Bird} {et~al.}(2022){Bird}, {Xue}, {Liu}, {Flynn}, {Shen}, {Wang},
  {Yang}, {Zhai}, {Zhu}, {Zhao}, \& {Tian}}]{2022MNRAS.516..731B}
{Bird}, S.~A., {Xue}, X.-X., {Liu}, C., {et~al.} 2022, \mnras, 516, 731,
  \dodoi{10.1093/mnras/stac2036}

\bibitem[{{Bovy} {et~al.}(2012){Bovy}, {Allende Prieto}, {Beers}, {Bizyaev},
  {da Costa}, {Cunha}, {Ebelke}, {Eisenstein}, {Frinchaboy}, {Garc{\'\i}a
  P{\'e}rez}, {Girardi}, {Hearty}, {Hogg}, {Holtzman}, {Maia}, {Majewski},
  {Malanushenko}, {Malanushenko}, {M{\'e}sz{\'a}ros}, {Nidever}, {O'Connell},
  {O'Donnell}, {Oravetz}, {Pan}, {Rocha-Pinto}, {Schiavon}, {Schneider},
  {Schultheis}, {Skrutskie}, {Smith}, {Weinberg}, {Wilson}, \&
  {Zasowski}}]{Bovy_etal_2012}
{Bovy}, J., {Allende Prieto}, C., {Beers}, T.~C., {et~al.} 2012, Astrophys.J.,
  759, 131, \dodoi{10.1088/0004-637X/759/2/131}

\bibitem[{{Chen} {et~al.}(2001){Chen}, {Stoughton}, {Smith}, {Uomoto}, {Pier},
  {Yanny}, {Ivezi{\'c}}, {York}, {Anderson}, {Annis}, {Brinkmann}, {Csabai},
  {Fukugita}, {Hindsley}, {Lupton}, {Munn}, \& {SDSS
  Collaboration}}]{Chen_etal_2001}
{Chen}, B., {Stoughton}, C., {Smith}, J.~A., {et~al.} 2001, Astrophys.J., 553,
  184, \dodoi{10.1086/320647}

\bibitem[{{Chrob{\'a}kov{\'a}} {et~al.}(2020){Chrob{\'a}kov{\'a}},
  {L{\'o}pez-Corredoira}, {Sylos Labini}, {Wang}, \&
  {Nagy}}]{Chrobakova_etal_2020}
{Chrob{\'a}kov{\'a}}, {\v{Z}}., {L{\'o}pez-Corredoira}, M., {Sylos Labini}, F.,
  {Wang}, H.~F., \& {Nagy}, R. 2020, Astron.Astrohys, 642, A95,
  \dodoi{10.1051/0004-6361/202038736}

\bibitem[{{Dias} \& {L{\'e}pine}(2005)}]{Dias+Lepine_2005}
{Dias}, W.~S., \& {L{\'e}pine}, J.~R.~D. 2005, Astrophys.J., 629, 825,
  \dodoi{10.1086/431456}

\bibitem[{{Drimmel} {et~al.}(2022){Drimmel}, {Romero-Gomez}, {Chemin}, {Ramos},
  {Poggio}, {Ripepi}, {Andrae}, {Blomme}, {Cantat-Gaudin}, {Castro-Ginard},
  {Clementini}, {Figueras}, {Fouesneau}, {Fremat}, {Jardine}, {Khanna},
  {Lobel}, {Marshall}, {Muraveva}, {Brown}, {Vallenari}, {Prusti}, {de
  Bruijne}, {Arenou}, {Babusiaux}, {Biermann}, {Creevey}, {Ducourant}, {Evans},
  {Eyer}, {Guerra}, {Hutton}, {Jordi}, {Klioner}, {Lammers}, {Lindegren},
  {Luri}, {Mignard}, {Panem}, {Pourbaix}, {Randich}, {Sartoretti}, {Soubiran},
  {Tanga}, {Walton}, {Bailer-Jones}, {Bastian}, {Jansen}, {Katz}, {Lattanzi},
  {van Leeuwen}, {Bakker}, {Cacciari}, {Casta{\~n}eda}, {De Angeli},
  {Fabricius}, {Galluccio}, {Guerrier}, {Heiter}, {Masana}, {Messineo},
  {Mowlavi}, {Nicolas}, {Nienartowicz}, {Pailler}, {Panuzzo}, {Riclet}, {Roux},
  {Seabroke}, {Sordo{\o}rcit}, {Th{\'e}venin}, {Gracia-Abril}, {Portell},
  {Teyssier}, {Altmann}, {Audard}, {Bellas-Velidis}, {Benson}, {Berthier},
  {Burgess}, {Busonero}, {Busso}, {C{\'a}novas}, {Carry}, {Cellino}, {Cheek},
  {Damerdji}, {Davidson}, {de Teodoro}, {Nu{\~n}ez Campos}, {Delchambre},
  {Dell'Oro}, {Esquej}, {Fern{\'a}ndez-Hern{\'a}ndez}, {Fraile}, {Garabato},
  {Garc{\'\i}a-Lario}, {Gosset}, {Haigron}, {Halbwachs}, {Hambly}, {Harrison},
  {Hern{\'a}ndez}, {Hestroffer}, {Hodgkin}, {Holl}, {Jan{\ss}en}, {Jevardat de
  Fombelle}, {Jordan}, {Krone-Martins}, {Lanzafame}, {L{\"o}ffler}, {Marchal},
  {Marrese}, {Moitinho}, {Muinonen}, {Osborne}, {Pancino}, {Pauwels},
  {Recio-Blanco}, {Reyl{\'e}}, {Riello}, {Rimoldini}, {Roegiers}, {Rybizki},
  {Sarro}, {Siopis}, {Smith}, {Sozzetti}, {Utrilla}, {van Leeuwen}, {Abbas},
  {{\'A}brah{\'a}m}, {Abreu Aramburu}, {Aerts}, {Aguado}, {Ajaj},
  {Aldea-Montero}, {Altavilla}, {{\'A}lvarez}, {Alves}, {Anders}, {Anderson},
  {Anglada Varela}, {Antoja}, {Baines}, {Baker}, {Balaguer-N{\'u}{\~n}ez},
  {Balbinot}, {Balog}, {Barache}, {Barbato}, {Barros}, {Barstow},
  {Bartolom{\'e}}, {Bassilana}, {Bauchet}, {Becciani}, {Bellazzini},
  {Berihuete}, {Bernet}, {Bertone}, {Bianchi}, {Binnenfeld}, {Blanco-Cuaresma},
  {Blazere}, {Boch}, {Bombrun}, {Bossini}, {Bouquillon}, {Bragaglia},
  {Bramante}, {Breedt}, {Bressan}, {Brouillet}, {Brugaletta}, {Bucciarelli},
  {Burlacu}, {Butkevich}, {Buzzi}, {Caffau}, {Cancelliere}, {Carballo},
  {Carlucci}, {Carnerero}, {Carrasco}, {Casamiquela}, {Castellani}, {Chaoul},
  {Charlot}, {Chiaramida}, {Chiavassa}, {Chornay}, {Comoretto}, {Contursi},
  {Cooper}, {Cornez}, {Cowell}, {Crifo}, {Cropper}, {Crosta}, {Crowley},
  {Dafonte}, {Dapergolas}, {David}, {David}, {de Laverny}, {De Luise}, {De
  March}, {De Ridder}, {de Souza}, {de Torres}, {del Peloso}, {del Pozo},
  {Delbo}, {Delgado}, {Delisle}, {Demouchy}, {Dharmawardena}, {Di Matteo},
  {Diakite}, {Diener}, {Distefano}, {Dolding}, {Edvardsson}, {Enke}, {Fabre},
  {Fabrizio}, {Faigler}, {Fedorets}, {Fernique}, {Fienga}, {Fournier},
  {Fouron}, {Fragkoudi}, {Gai}, {Garcia-Gutierrez}, {Garcia-Reinaldos},
  {Garc{\'\i}a-Torres}, {Garofalo}, {Gavel}, {Gavras}, {Gerlach}, {Geyer},
  {Giacobbe}, {Gilmore}, {Girona}, {Giuffrida}, {Gomel}, {Gomez},
  {Gonz{\'a}lez-N{\'u}{\~n}ez}, {Gonz{\'a}lez-Santamar{\'\i}a},
  {Gonz{\'a}lez-Vidal}, {Granvik}, {Guillout}, {Guiraud},
  {Guti{\'e}rrez-S{\'a}nchez}, {Guy}, {Hatzidimitriou}, {Hauser}, {Haywood},
  {Helmer}, {Helmi}, {Sarmiento}, {Hidalgo}, {Hilger}, {H{\l}adczuk}, {Hobbs},
  {Holland}, {Huckle}, {Jasniewicz}, {Jean-Antoine Piccolo},
  {Jim{\'e}nez-Arranz}, {Jorissen}, {Juaristi Campillo}, {Julbe}, {Karbevska},
  {Kervella}, {Kontizas}, {Kordopatis}, {Korn}, {K{\'o}sp{\'a}l},
  {Kostrzewa-Rutkowska}, {Kruszy{\'n}ska}, {Kun}, {Laizeau}, {Lambert},
  {Lanza}, {Lasne}, {Le Campion}, {Lebreton}, {Lebzelter}, {Leccia}, {Leclerc},
  {Lecoeur-Taibi}, {Liao}, {Licata}, {Lindstr{\o}m}, {Lister}, {Livanou},
  {Lorca}, {Loup}, {Madrero Pardo}, {Magdaleno Romeo}, {Managau}, {Mann},
  {Manteiga}, {Marchant}, {Marconi}, {Marcos}, {Marcos Santos}, {Mar{\'\i}n
  Pina}, {Marinoni}, {Marocco}, {Polo}, {Mart{\'\i}n-Fleitas}, {Marton},
  {Mary}, {Masip}, {Massari}, {Mastrobuono-Battisti}, {Mazeh}, {McMillan},
  {Messina}, {Michalik}, {Millar}, {Mints}, {Molina}, {Molinaro}, {Moln{\'a}r},
  {Monari}, {Mongui{\'o}}, {Montegriffo}, {Montero}, {Mor}, {Mora},
  {Morbidelli}, {Morel}, {Morris}, {Murphy}, {Musella}, {Nagy}, {Noval},
  {Oca{\~n}a}, {Ogden}, {Ordenovic}, {Osinde}, {Pagani}, {Pagano}, {Palaversa},
  {Palicio}, {Pallas-Quintela}, {Panahi}, {Payne-Wardenaar}, {Pe{\~n}alosa
  Esteller}, {Penttil{\"a}}, {Pichon}, {Piersimoni}, {Pineau}, {Plachy},
  {Plum}, {Pr{\v{s}}a}, {Pulone}, {Racero}, {Ragaini}, {Rainer}, {Raiteri},
  {Rambaux}, {Ramos-Lerate}, {Re Fiorentin}, {Regibo}, {Richards}, {Rios Diaz},
  {Riva}, {Rix}, {Rixon}, {Robichon}, {Robin}, {Robin}, {Roelens}, {Rogues},
  {Rohrbasser}, {Rowell}, {Royer}, {Ruz Mieres}, {Rybicki}, {Sadowski},
  {S{\'a}ez N{\'u}{\~n}ez}, {Sagrist{\`a} Sell{\'e}s}, {Sahlmann}, {Salguero},
  {Samaras}, {Sanchez Gimenez}, {Sanna}, {Santove{\~n}a}, {Sarasso},
  {Schultheis}, {Sciacca}, {Segol}, {Segovia}, {S{\'e}gransan}, {Semeux},
  {Shahaf}, {Siddiqui}, {Siebert}, {Siltala}, {Silvelo}, {Slezak}, {Slezak},
  {Smart}, {Snaith}, {Solano}, {Solitro}, {Souami}, {Souchay}, {Spagna},
  {Spina}, {Spoto}, {Steele}, {Steidelm{\"u}ller}, {Stephenson}, {S{\"u}veges},
  {Surdej}, {Szabados}, {Szegedi-Elek}, {Taris}, {Taylo}, {Teixeira},
  {Tolomei}, {Tonello}, {Torra}, {Torra}, {Torralba Elipe}, {Trabucchi},
  {Tsounis}, {Turon}, {Ulla}, {Unger}, {Vaillant}, {van Dillen}, {van Reeven},
  {Vanel}, {Vecchiato}, {Viala}, {Vicente}, {Voutsinas}, {Weiler}, {Wevers},
  {Wyrzykowski}, {Yoldas}, {Yvard}, {Zhao}, {Zorec}, {Zucker}, \&
  {Zwitter}}]{Drimmel2022}
{Drimmel}, R., {Romero-Gomez}, M., {Chemin}, L., {et~al.} 2022, arXiv e-prints,
  arXiv:2206.06207.
\newblock \doarXiv{2206.06207}

\bibitem[{{Eilers} {et~al.}(2019){Eilers}, {Hogg}, {Rix}, \&
  {Ness}}]{Eilers_etal_2019}
{Eilers}, A.-C., {Hogg}, D.~W., {Rix}, H.-W., \& {Ness}, M.~K. 2019,
  Astrophys.J., 871, 120, \dodoi{10.3847/1538-4357/aaf648}

\bibitem[{{Gaia Collaboration} {et~al.}(2016){Gaia Collaboration}, {Prusti},
  {de Bruijne}, {Brown}, {Vallenari}, {Babusiaux}, {Bailer-Jones}, {Bastian},
  {Biermann}, {Evans}, \& et~al.}]{Gaia_2016}
{Gaia Collaboration}, {Prusti}, T., {de Bruijne}, J.~H.~J., {et~al.} 2016,
  Astron.Astrophys., 595, A1, \dodoi{10.1051/0004-6361/201629272}

\bibitem[{{Gaia Collaboration} {et~al.}(2018){Gaia Collaboration}, {Katz},
  {Antoja}, {Romero-G{\'o}mez}, {Drimmel}, {Reyl{\'e}}, {Seabroke}, {Soubiran},
  {Babusiaux}, {Di Matteo}, \& et~al.}]{Katz_etal_2018}
{Gaia Collaboration}, {Katz}, D., {Antoja}, T., {et~al.} 2018,
  Astron.Astrophys., 616, A11, \dodoi{10.1051/0004-6361/201832865}

\bibitem[{{Gaia Collaboration} {et~al.}(2021){Gaia Collaboration}, {Antoja},
  {McMillan}, {Kordopatis}, {Ramos}, {Helmi}, {Balbinot}, {Cantat-Gaudin},
  {Chemin}, {Figueras}, {Jordi}, {Khanna}, {Romero-G{\'o}mez}, {Seabroke},
  {Brown}, {Vallenari}, {Prusti}, {de Bruijne}, {Babusiaux}, {Biermann},
  {Creevey}, {Evans}, {Eyer}, {Hutton}, {Jansen}, {Klioner}, {Lammers},
  {Lindegren}, {Luri}, {Mignard}, {Panem}, {Pourbaix}, {Randich}, {Sartoretti},
  {Soubiran}, {Walton}, {Arenou}, {Bailer-Jones}, {Bastian}, {Cropper},
  {Drimmel}, {Katz}, {Lattanzi}, {van Leeuwen}, {Bakker}, {Casta{\~n}eda}, {De
  Angeli}, {Ducourant}, {Fabricius}, {Fouesneau}, {Fr{\'e}mat}, {Guerra},
  {Guerrier}, {Guiraud}, {Jean-Antoine Piccolo}, {Masana}, {Messineo},
  {Mowlavi}, {Nicolas}, {Nienartowicz}, {Pailler}, {Panuzzo}, {Riclet}, {Roux},
  {Sordo}, {Tanga}, {Th{\'e}venin}, {Gracia-Abril}, {Portell}, {Teyssier},
  {Altmann}, {Andrae}, {Bellas-Velidis}, {Benson}, {Berthier}, {Blomme},
  {Brugaletta}, {Burgess}, {Busso}, {Carry}, {Cellino}, {Cheek}, {Clementini},
  {Damerdji}, {Davidson}, {Delchambre}, {Dell'Oro},
  {Fern{\'a}ndez-Hern{\'a}ndez}, {Galluccio}, {Garc{\'\i}a-Lario},
  {Garcia-Reinaldos}, {Gonz{\'a}lez-N{\'u}{\~n}ez}, {Gosset}, {Haigron},
  {Halbwachs}, {Hambly}, {Harrison}, {Hatzidimitriou}, {Heiter},
  {Hern{\'a}ndez}, {Hestroffer}, {Hodgkin}, {Holl}, {Jan{\ss}en}, {Jevardat de
  Fombelle}, {Jordan}, {Krone-Martins}, {Lanzafame}, {L{\"o}ffler}, {Lorca},
  {Manteiga}, {Marchal}, {Marrese}, {Moitinho}, {Mora}, {Muinonen}, {Osborne},
  {Pancino}, {Pauwels}, {Recio-Blanco}, {Richards}, {Riello}, {Rimoldini},
  {Robin}, {Roegiers}, {Rybizki}, {Sarro}, {Siopis}, {Smith}, {Sozzetti},
  {Ulla}, {Utrilla}, {van Leeuwen}, {van Reeven}, {Abbas}, {Abreu Aramburu},
  {Accart}, {Aerts}, {Aguado}, {Ajaj}, {Altavilla}, {{\'A}lvarez}, {{\'A}lvarez
  Cid-Fuentes}, {Alves}, {Anderson}, {Varela}, {Audard}, {Baines}, {Baker},
  {Balaguer-N{\'u}{\~n}ez}, {Balog}, {Barache}, {Barbato}, {Barros}, {Barstow},
  {Bartolom{\'e}}, {Bassilana}, {Bauchet}, {Baudesson-Stella}, {Becciani},
  {Bellazzini}, {Bernet}, {Bertone}, {Bianchi}, {Blanco-Cuaresma}, {Boch},
  {Bombrun}, {Bossini}, {Bouquillon}, {Bragaglia}, {Bramante}, {Breedt},
  {Bressan}, {Brouillet}, {Bucciarelli}, {Burlacu}, {Busonero}, {Butkevich},
  {Buzzi}, {Caffau}, {Cancelliere}, {C{\'a}novas}, {Carballo}, {Carlucci},
  {Carnerero}, {Carrasco}, {Casamiquela}, {Castellani}, {Castro-Ginard},
  {Castro Sampol}, {Chaoul}, {Charlot}, {Chiavassa}, {Cioni}, {Comoretto},
  {Cooper}, {Cornez}, {Cowell}, {Crifo}, {Crosta}, {Crowley}, {Dafonte},
  {Dapergolas}, {David}, {David}, {de Laverny}, {De Luise}, {De March}, {De
  Ridder}, {de Souza}, {de Teodoro}, {de Torres}, {del Peloso}, {del Pozo},
  {Delgado}, {Delgado}, {Delisle}, {Di Matteo}, {Diakite}, {Diener},
  {Distefano}, {Dolding}, {Eappachen}, {Enke}, {Esquej}, {Fabre}, {Fabrizio},
  {Faigler}, {Fedorets}, {Fernique}, {Fienga}, {Fouron}, {Fragkoudi}, {Fraile},
  {Franke}, {Gai}, {Garabato}, {Garcia-Gutierrez}, {Garc{\'\i}a-Torres},
  {Garofalo}, {Gavras}, {Gerlach}, {Geyer}, {Giacobbe}, {Gilmore}, {Girona},
  {Giuffrida}, {Gomez}, {Gonzalez-Santamaria}, {Gonz{\'a}lez-Vidal}, {Granvik},
  {Guti{\'e}rrez-S{\'a}nchez}, {Guy}, {Hauser}, {Haywood}, {Hidalgo}, {Hilger},
  {H{\l}adczuk}, {Hobbs}, {Holland}, {Huckle}, {Jasniewicz}, {Jonker},
  {Juaristi Campillo}, {Julbe}, {Karbevska}, {Kervella}, {Kochoska},
  {Kontizas}, {Korn}, {Kostrzewa-Rutkowska}, {Kruszy{\'n}ska}, {Lambert},
  {Lanza}, {Lasne}, {Le Campion}, {Le Fustec}, {Lebreton}, {Lebzelter},
  {Leccia}, {Leclerc}, {Lecoeur-Taibi}, {Liao}, {Licata}, {Lindstr{\o}m},
  {Lister}, {Livanou}, {Lobel}, {Madrero Pardo}, {Managau}, {Mann}, {Marchant},
  {Marconi}, {Marcos Santos}, {Marinoni}, {Marocco}, {Marshall}, {Martin Polo},
  {Mart{\'\i}n-Fleitas}, {Masip}, {Massari}, {Mastrobuono-Battisti}, {Mazeh},
  {Messina}, {Michalik}, {Millar}, {Mints}, {Molina}, {Molinaro}, {Moln{\'a}r},
  {Montegriffo}, {Mor}, {Morbidelli}, {Morel}, {Morris}, {Mulone}, {Munoz},
  {Muraveva}, {Murphy}, {Musella}, {Noval}, {Ord{\'e}novic}, {Orr{\`u}},
  {Osinde}, {Pagani}, {Pagano}, {Palaversa}, {Palicio}, {Panahi}, {Pawlak},
  {Pe{\~n}alosa Esteller}, {Penttil{\"a}}, {Piersimoni}, {Pineau}, {Plachy},
  {Plum}, {Poggio}, {Poretti}, {Poujoulet}, {Pr{\v{s}}a}, {Pulone}, {Racero},
  {Ragaini}, {Rainer}, {Raiteri}, {Rambaux}, {Ramos-Lerate}, {Re Fiorentin},
  {Regibo}, {Reyl{\'e}}, {Ripepi}, {Riva}, {Rixon}, {Robichon}, {Robin},
  {Roelens}, {Rohrbasser}, {Rowell}, {Royer}, {Rybicki}, {Sadowski},
  {Sagrist{\`a} Sell{\'e}s}, {Sahlmann}, {Salgado}, {Salguero}, {Samaras},
  {Sanchez Gimenez}, {Sanna}, {Santove{\~n}a}, {Sarasso}, {Schultheis},
  {Sciacca}, {Segol}, {Segovia}, {S{\'e}gransan}, {Semeux}, {Siddiqui},
  {Siebert}, {Siltala}, {Slezak}, {Smart}, {Solano}, {Solitro}, {Souami},
  {Souchay}, {Spagna}, {Spoto}, {Steele}, {Steidelm{\"u}ller}, {Stephenson},
  {S{\"u}veges}, {Szabados}, {Szegedi-Elek}, {Taris}, {Tauran}, {Taylor},
  {Teixeira}, {Thuillot}, {Tonello}, {Torra}, {Torra}, {Turon}, {Unger},
  {Vaillant}, {van Dillen}, {Vanel}, {Vecchiato}, {Viala}, {Vicente},
  {Voutsinas}, {Weiler}, {Wevers}, {Wyrzykowski}, {Yoldas}, {Yvard}, {Zhao},
  {Zorec}, {Zucker}, {Zurbach}, \& {Zwitter}}]{2021antoja}
{Gaia Collaboration}, {Antoja}, T., {McMillan}, P.~J., {et~al.} 2021, \aap,
  649, A8, \dodoi{10.1051/0004-6361/202039714}

\bibitem[{{Gaia Collaboration} {et~al.}(2022{\natexlab{a}}){Gaia
  Collaboration}, {Vallenari}, {Brown}, {Prusti}, {de Bruijne}, {Arenou},
  {Babusiaux}, {Biermann}, {Creevey}, {Ducourant}, {Evans}, {Eyer}, {Guerra},
  {Hutton}, {Jordi}, {Klioner}, {Lammers}, {Lindegren}, {Luri}, {Mignard},
  {Panem}, {Pourbaix}, {Randich}, {Sartoretti}, {Soubiran}, {Tanga}, {Walton},
  {Bailer-Jones}, {Bastian}, {Drimmel}, {Jansen}, {Katz}, {Lattanzi}, {van
  Leeuwen}, {Bakker}, {Cacciari}, {Casta{\~n}eda}, {De Angeli}, {Fabricius},
  {Fouesneau}, {Fr{\'e}mat}, {Galluccio}, {Guerrier}, {Heiter}, {Masana},
  {Messineo}, {Mowlavi}, {Nicolas}, {Nienartowicz}, {Pailler}, {Panuzzo},
  {Riclet}, {Roux}, {Seabroke}, {Sordo{\o}rcit}, {Th{\'e}venin},
  {Gracia-Abril}, {Portell}, {Teyssier}, {Altmann}, {Andrae}, {Audard},
  {Bellas-Velidis}, {Benson}, {Berthier}, {Blomme}, {Burgess}, {Busonero},
  {Busso}, {C{\'a}novas}, {Carry}, {Cellino}, {Cheek}, {Clementini},
  {Damerdji}, {Davidson}, {de Teodoro}, {Nu{\~n}ez Campos}, {Delchambre},
  {Dell'Oro}, {Esquej}, {Fern{\'a}ndez-Hern{\'a}ndez}, {Fraile}, {Garabato},
  {Garc{\'\i}a-Lario}, {Gosset}, {Haigron}, {Halbwachs}, {Hambly}, {Harrison},
  {Hern{\'a}ndez}, {Hestroffer}, {Hodgkin}, {Holl}, {Jan{\ss}en}, {Jevardat de
  Fombelle}, {Jordan}, {Krone-Martins}, {Lanzafame}, {L{\"o}ffler}, {Marchal},
  {Marrese}, {Moitinho}, {Muinonen}, {Osborne}, {Pancino}, {Pauwels},
  {Recio-Blanco}, {Reyl{\'e}}, {Riello}, {Rimoldini}, {Roegiers}, {Rybizki},
  {Sarro}, {Siopis}, {Smith}, {Sozzetti}, {Utrilla}, {van Leeuwen}, {Abbas},
  {{\'A}brah{\'a}m}, {Abreu Aramburu}, {Aerts}, {Aguado}, {Ajaj},
  {Aldea-Montero}, {Altavilla}, {{\'A}lvarez}, {Alves}, {Anders}, {Anderson},
  {Anglada Varela}, {Antoja}, {Baines}, {Baker}, {Balaguer-N{\'u}{\~n}ez},
  {Balbinot}, {Balog}, {Barache}, {Barbato}, {Barros}, {Barstow},
  {Bartolom{\'e}}, {Bassilana}, {Bauchet}, {Becciani}, {Bellazzini},
  {Berihuete}, {Bernet}, {Bertone}, {Bianchi}, {Binnenfeld}, {Blanco-Cuaresma},
  {Blazere}, {Boch}, {Bombrun}, {Bossini}, {Bouquillon}, {Bragaglia},
  {Bramante}, {Breedt}, {Bressan}, {Brouillet}, {Brugaletta}, {Bucciarelli},
  {Burlacu}, {Butkevich}, {Buzzi}, {Caffau}, {Cancelliere}, {Cantat-Gaudin},
  {Carballo}, {Carlucci}, {Carnerero}, {Carrasco}, {Casamiquela}, {Castellani},
  {Castro-Ginard}, {Chaoul}, {Charlot}, {Chemin}, {Chiaramida}, {Chiavassa},
  {Chornay}, {Comoretto}, {Contursi}, {Cooper}, {Cornez}, {Cowell}, {Crifo},
  {Cropper}, {Crosta}, {Crowley}, {Dafonte}, {Dapergolas}, {David}, {David},
  {de Laverny}, {De Luise}, {De March}, {De Ridder}, {de Souza}, {de Torres},
  {del Peloso}, {del Pozo}, {Delbo}, {Delgado}, {Delisle}, {Demouchy},
  {Dharmawardena}, {Di Matteo}, {Diakite}, {Diener}, {Distefano}, {Dolding},
  {Edvardsson}, {Enke}, {Fabre}, {Fabrizio}, {Faigler}, {Fedorets}, {Fernique},
  {Fienga}, {Figueras}, {Fournier}, {Fouron}, {Fragkoudi}, {Gai},
  {Garcia-Gutierrez}, {Garcia-Reinaldos}, {Garc{\'\i}a-Torres}, {Garofalo},
  {Gavel}, {Gavras}, {Gerlach}, {Geyer}, {Giacobbe}, {Gilmore}, {Girona},
  {Giuffrida}, {Gomel}, {Gomez}, {Gonz{\'a}lez-N{\'u}{\~n}ez},
  {Gonz{\'a}lez-Santamar{\'\i}a}, {Gonz{\'a}lez-Vidal}, {Granvik}, {Guillout},
  {Guiraud}, {Guti{\'e}rrez-S{\'a}nchez}, {Guy}, {Hatzidimitriou}, {Hauser},
  {Haywood}, {Helmer}, {Helmi}, {Sarmiento}, {Hidalgo}, {Hilger},
  {H{\l}adczuk}, {Hobbs}, {Holland}, {Huckle}, {Jardine}, {Jasniewicz},
  {Jean-Antoine Piccolo}, {Jim{\'e}nez-Arranz}, {Jorissen}, {Juaristi
  Campillo}, {Julbe}, {Karbevska}, {Kervella}, {Khanna}, {Kontizas},
  {Kordopatis}, {Korn}, {K{\'o}sp{\'a}l}, {Kostrzewa-Rutkowska},
  {Kruszy{\'n}ska}, {Kun}, {Laizeau}, {Lambert}, {Lanza}, {Lasne}, {Le
  Campion}, {Lebreton}, {Lebzelter}, {Leccia}, {Leclerc}, {Lecoeur-Taibi},
  {Liao}, {Licata}, {Lindstr{\o}m}, {Lister}, {Livanou}, {Lobel}, {Lorca},
  {Loup}, {Madrero Pardo}, {Magdaleno Romeo}, {Managau}, {Mann}, {Manteiga},
  {Marchant}, {Marconi}, {Marcos}, {Marcos Santos}, {Mar{\'\i}n Pina},
  {Marinoni}, {Marocco}, {Marshall}, {Polo}, {Mart{\'\i}n-Fleitas}, {Marton},
  {Mary}, {Masip}, {Massari}, {Mastrobuono-Battisti}, {Mazeh}, {McMillan},
  {Messina}, {Michalik}, {Millar}, {Mints}, {Molina}, {Molinaro}, {Moln{\'a}r},
  {Monari}, {Mongui{\'o}}, {Montegriffo}, {Montero}, {Mor}, {Mora},
  {Morbidelli}, {Morel}, {Morris}, {Muraveva}, {Murphy}, {Musella}, {Nagy},
  {Noval}, {Oca{\~n}a}, {Ogden}, {Ordenovic}, {Osinde}, {Pagani}, {Pagano},
  {Palaversa}, {Palicio}, {Pallas-Quintela}, {Panahi}, {Payne-Wardenaar},
  {Pe{\~n}alosa Esteller}, {Penttil{\"a}}, {Pichon}, {Piersimoni}, {Pineau},
  {Plachy}, {Plum}, {Poggio}, {Pr{\v{s}}a}, {Pulone}, {Racero}, {Ragaini},
  {Rainer}, {Raiteri}, {Rambaux}, {Ramos}, {Ramos-Lerate}, {Re Fiorentin},
  {Regibo}, {Richards}, {Rios Diaz}, {Ripepi}, {Riva}, {Rix}, {Rixon},
  {Robichon}, {Robin}, {Robin}, {Roelens}, {Rogues}, {Rohrbasser},
  {Romero-G{\'o}mez}, {Rowell}, {Royer}, {Ruz Mieres}, {Rybicki}, {Sadowski},
  {S{\'a}ez N{\'u}{\~n}ez}, {Sagrist{\`a} Sell{\'e}s}, {Sahlmann}, {Salguero},
  {Samaras}, {Sanchez Gimenez}, {Sanna}, {Santove{\~n}a}, {Sarasso},
  {Schultheis}, {Sciacca}, {Segol}, {Segovia}, {S{\'e}gransan}, {Semeux},
  {Shahaf}, {Siddiqui}, {Siebert}, {Siltala}, {Silvelo}, {Slezak}, {Slezak},
  {Smart}, {Snaith}, {Solano}, {Solitro}, {Souami}, {Souchay}, {Spagna},
  {Spina}, {Spoto}, {Steele}, {Steidelm{\"u}ller}, {Stephenson}, {S{\"u}veges},
  {Surdej}, {Szabados}, {Szegedi-Elek}, {Taris}, {Taylo}, {Teixeira},
  {Tolomei}, {Tonello}, {Torra}, {Torra}, {Torralba Elipe}, {Trabucchi},
  {Tsounis}, {Turon}, {Ulla}, {Unger}, {Vaillant}, {van Dillen}, {van Reeven},
  {Vanel}, {Vecchiato}, {Viala}, {Vicente}, {Voutsinas}, {Weiler}, {Wevers},
  {Wyrzykowski}, {Yoldas}, {Yvard}, {Zhao}, {Zorec}, {Zucker}, \&
  {Zwitter}}]{Gaia_DR3}
{Gaia Collaboration}, {Vallenari}, A., {Brown}, A.~G.~A., {et~al.}
  2022{\natexlab{a}}, arXiv e-prints, arXiv:2208.00211.
\newblock \doarXiv{2208.00211}

\bibitem[{{Gaia Collaboration} {et~al.}(2022{\natexlab{b}}){Gaia
  Collaboration}, {Drimmel}, {Romero-Gomez}, {Chemin}, {Ramos}, {Poggio},
  {Ripepi}, {Andrae}, {Blomme}, {Cantat-Gaudin}, {Castro-Ginard}, {Clementini},
  {Figueras}, {Fouesneau}, {Fremat}, {Jardine}, {Khanna}, {Lobel}, {Marshall},
  {Muraveva}, {Brown}, {Vallenari}, {Prusti}, {de Bruijne}, {Arenou},
  {Babusiaux}, {Biermann}, {Creevey}, {Ducourant}, {Evans}, {Eyer}, {Guerra},
  {Hutton}, {Jordi}, {Klioner}, {Lammers}, {Lindegren}, {Luri}, {Mignard},
  {Panem}, {Pourbaix}, {Randich}, {Sartoretti}, {Soubiran}, {Tanga}, {Walton},
  {Bailer-Jones}, {Bastian}, {Jansen}, {Katz}, {Lattanzi}, {van Leeuwen},
  {Bakker}, {Cacciari}, {Casta{\~n}eda}, {De Angeli}, {Fabricius}, {Galluccio},
  {Guerrier}, {Heiter}, {Masana}, {Messineo}, {Mowlavi}, {Nicolas},
  {Nienartowicz}, {Pailler}, {Panuzzo}, {Riclet}, {Roux}, {Seabroke},
  {Sordo{\o}rcit}, {Th{\'e}venin}, {Gracia-Abril}, {Portell}, {Teyssier},
  {Altmann}, {Audard}, {Bellas-Velidis}, {Benson}, {Berthier}, {Burgess},
  {Busonero}, {Busso}, {C{\'a}novas}, {Carry}, {Cellino}, {Cheek}, {Damerdji},
  {Davidson}, {de Teodoro}, {Nu{\~n}ez Campos}, {Delchambre}, {Dell'Oro},
  {Esquej}, {Fern{\'a}ndez-Hern{\'a}ndez}, {Fraile}, {Garabato},
  {Garc{\'\i}a-Lario}, {Gosset}, {Haigron}, {Halbwachs}, {Hambly}, {Harrison},
  {Hern{\'a}ndez}, {Hestroffer}, {Hodgkin}, {Holl}, {Jan{\ss}en}, {Jevardat de
  Fombelle}, {Jordan}, {Krone-Martins}, {Lanzafame}, {L{\"o}ffler}, {Marchal},
  {Marrese}, {Moitinho}, {Muinonen}, {Osborne}, {Pancino}, {Pauwels},
  {Recio-Blanco}, {Reyl{\'e}}, {Riello}, {Rimoldini}, {Roegiers}, {Rybizki},
  {Sarro}, {Siopis}, {Smith}, {Sozzetti}, {Utrilla}, {van Leeuwen}, {Abbas},
  {{\'A}brah{\'a}m}, {Abreu Aramburu}, {Aerts}, {Aguado}, {Ajaj},
  {Aldea-Montero}, {Altavilla}, {{\'A}lvarez}, {Alves}, {Anders}, {Anderson},
  {Anglada Varela}, {Antoja}, {Baines}, {Baker}, {Balaguer-N{\'u}{\~n}ez},
  {Balbinot}, {Balog}, {Barache}, {Barbato}, {Barros}, {Barstow},
  {Bartolom{\'e}}, {Bassilana}, {Bauchet}, {Becciani}, {Bellazzini},
  {Berihuete}, {Bernet}, {Bertone}, {Bianchi}, {Binnenfeld}, {Blanco-Cuaresma},
  {Blazere}, {Boch}, {Bombrun}, {Bossini}, {Bouquillon}, {Bragaglia},
  {Bramante}, {Breedt}, {Bressan}, {Brouillet}, {Brugaletta}, {Bucciarelli},
  {Burlacu}, {Butkevich}, {Buzzi}, {Caffau}, {Cancelliere}, {Carballo},
  {Carlucci}, {Carnerero}, {Carrasco}, {Casamiquela}, {Castellani}, {Chaoul},
  {Charlot}, {Chiaramida}, {Chiavassa}, {Chornay}, {Comoretto}, {Contursi},
  {Cooper}, {Cornez}, {Cowell}, {Crifo}, {Cropper}, {Crosta}, {Crowley},
  {Dafonte}, {Dapergolas}, {David}, {David}, {de Laverny}, {De Luise}, {De
  March}, {De Ridder}, {de Souza}, {de Torres}, {del Peloso}, {del Pozo},
  {Delbo}, {Delgado}, {Delisle}, {Demouchy}, {Dharmawardena}, {Di Matteo},
  {Diakite}, {Diener}, {Distefano}, {Dolding}, {Edvardsson}, {Enke}, {Fabre},
  {Fabrizio}, {Faigler}, {Fedorets}, {Fernique}, {Fienga}, {Fournier},
  {Fouron}, {Fragkoudi}, {Gai}, {Garcia-Gutierrez}, {Garcia-Reinaldos},
  {Garc{\'\i}a-Torres}, {Garofalo}, {Gavel}, {Gavras}, {Gerlach}, {Geyer},
  {Giacobbe}, {Gilmore}, {Girona}, {Giuffrida}, {Gomel}, {Gomez},
  {Gonz{\'a}lez-N{\'u}{\~n}ez}, {Gonz{\'a}lez-Santamar{\'\i}a},
  {Gonz{\'a}lez-Vidal}, {Granvik}, {Guillout}, {Guiraud},
  {Guti{\'e}rrez-S{\'a}nchez}, {Guy}, {Hatzidimitriou}, {Hauser}, {Haywood},
  {Helmer}, {Helmi}, {Sarmiento}, {Hidalgo}, {Hilger}, {H{\l}adczuk}, {Hobbs},
  {Holland}, {Huckle}, {Jasniewicz}, {Jean-Antoine Piccolo},
  {Jim{\'e}nez-Arranz}, {Jorissen}, {Juaristi Campillo}, {Julbe}, {Karbevska},
  {Kervella}, {Kontizas}, {Kordopatis}, {Korn}, {K{\'o}sp{\'a}l},
  {Kostrzewa-Rutkowska}, {Kruszy{\'n}ska}, {Kun}, {Laizeau}, {Lambert},
  {Lanza}, {Lasne}, {Le Campion}, {Lebreton}, {Lebzelter}, {Leccia}, {Leclerc},
  {Lecoeur-Taibi}, {Liao}, {Licata}, {Lindstr{\o}m}, {Lister}, {Livanou},
  {Lorca}, {Loup}, {Madrero Pardo}, {Magdaleno Romeo}, {Managau}, {Mann},
  {Manteiga}, {Marchant}, {Marconi}, {Marcos}, {Marcos Santos}, {Mar{\'\i}n
  Pina}, {Marinoni}, {Marocco}, {Polo}, {Mart{\'\i}n-Fleitas}, {Marton},
  {Mary}, {Masip}, {Massari}, {Mastrobuono-Battisti}, {Mazeh}, {McMillan},
  {Messina}, {Michalik}, {Millar}, {Mints}, {Molina}, {Molinaro}, {Moln{\'a}r},
  {Monari}, {Mongui{\'o}}, {Montegriffo}, {Montero}, {Mor}, {Mora},
  {Morbidelli}, {Morel}, {Morris}, {Murphy}, {Musella}, {Nagy}, {Noval},
  {Oca{\~n}a}, {Ogden}, {Ordenovic}, {Osinde}, {Pagani}, {Pagano}, {Palaversa},
  {Palicio}, {Pallas-Quintela}, {Panahi}, {Payne-Wardenaar}, {Pe{\~n}alosa
  Esteller}, {Penttil{\"a}}, {Pichon}, {Piersimoni}, {Pineau}, {Plachy},
  {Plum}, {Pr{\v{s}}a}, {Pulone}, {Racero}, {Ragaini}, {Rainer}, {Raiteri},
  {Rambaux}, {Ramos-Lerate}, {Re Fiorentin}, {Regibo}, {Richards}, {Rios Diaz},
  {Riva}, {Rix}, {Rixon}, {Robichon}, {Robin}, {Robin}, {Roelens}, {Rogues},
  {Rohrbasser}, {Rowell}, {Royer}, {Ruz Mieres}, {Rybicki}, {Sadowski},
  {S{\'a}ez N{\'u}{\~n}ez}, {Sagrist{\`a} Sell{\'e}s}, {Sahlmann}, {Salguero},
  {Samaras}, {Sanchez Gimenez}, {Sanna}, {Santove{\~n}a}, {Sarasso},
  {Schultheis}, {Sciacca}, {Segol}, {Segovia}, {S{\'e}gransan}, {Semeux},
  {Shahaf}, {Siddiqui}, {Siebert}, {Siltala}, {Silvelo}, {Slezak}, {Slezak},
  {Smart}, {Snaith}, {Solano}, {Solitro}, {Souami}, {Souchay}, {Spagna},
  {Spina}, {Spoto}, {Steele}, {Steidelm{\"u}ller}, {Stephenson}, {S{\"u}veges},
  {Surdej}, {Szabados}, {Szegedi-Elek}, {Taris}, {Taylo}, {Teixeira},
  {Tolomei}, {Tonello}, {Torra}, {Torra}, {Torralba Elipe}, {Trabucchi},
  {Tsounis}, {Turon}, {Ulla}, {Unger}, {Vaillant}, {van Dillen}, {van Reeven},
  {Vanel}, {Vecchiato}, {Viala}, {Vicente}, {Voutsinas}, {Weiler}, {Wevers},
  {Wyrzykowski}, {Yoldas}, {Yvard}, {Zhao}, {Zorec}, {Zucker}, \&
  {Zwitter}}]{Gaia_DR3_K}
{Gaia Collaboration}, {Drimmel}, R., {Romero-Gomez}, M., {et~al.}
  2022{\natexlab{b}}, arXiv e-prints, arXiv:2206.06207.
\newblock \doarXiv{2206.06207}

\bibitem[{{Galazutdinov} {et~al.}(2015){Galazutdinov}, {Strobel}, {Musaev},
  {Bondar}, \& {Kre{\l}owski}}]{GalazutdinoV_etal_2015}
{Galazutdinov}, G., {Strobel}, A., {Musaev}, F.~A., {Bondar}, A., \&
  {Kre{\l}owski}, J. 2015, Pub.Astron.Soc.Pacific, 127, 126,
  \dodoi{10.1086/680211}

\bibitem[{{Gibbons} {et~al.}(2014){Gibbons}, {Belokurov}, \&
  {Evans}}]{2014MNRAS.445.3788G}
{Gibbons}, S.~L.~J., {Belokurov}, V., \& {Evans}, N.~W. 2014, \mnras, 445,
  3788, \dodoi{10.1093/mnras/stu1986}

\bibitem[{{Huang} {et~al.}(2016){Huang}, {Liu}, {Yuan}, {Xiang}, {Zhang},
  {Chen}, {Ren}, {Wang}, {Zhang}, {Hou}, {Wang}, \& {Cao}}]{Huang_etal_2016}
{Huang}, Y., {Liu}, X.~W., {Yuan}, H.~B., {et~al.} 2016, Mon.Not.R.Astr.Soc.,
  463, 2623, \dodoi{10.1093/mnras/stw2096}

\bibitem[{{Jiao} {et~al.}(2021){Jiao}, {Hammer}, {Wang}, \&
  {Yang}}]{2021A&A...654A..25J}
{Jiao}, Y., {Hammer}, F., {Wang}, J.~L., \& {Yang}, Y.~B. 2021, \aap, 654, A25,
  \dodoi{10.1051/0004-6361/202141058}

\bibitem[{{Kafle} {et~al.}(2012){Kafle}, {Sharma}, {Lewis}, \&
  {Bland-Hawthorn}}]{Kafle_etal_2012}
{Kafle}, P.~R., {Sharma}, S., {Lewis}, G.~F., \& {Bland-Hawthorn}, J. 2012,
  Astrophys.J., 761, 98, \dodoi{10.1088/0004-637X/761/2/98}

\bibitem[{{Katz} {et~al.}(2022){Katz}, {Sartoretti}, {Guerrier}, {Panuzzo},
  {Seabroke}, {Th{\'e}venin}, {Cropper}, {Benson}, {Blomme}, {Haigron},
  {Marchal}, {Smith}, {Baker}, {Chemin}, {Damerdji}, {David}, {Dolding},
  {Fr{\'e}mat}, {Gosset}, {Jan{\ss}en}, {Jasniewicz}, {Lobel}, {Plum},
  {Samaras}, {Snaith}, {Soubiran}, {Vanel}, {Zwitter}, {Antoja}, {Arenou},
  {Babusiaux}, {Brouillet}, {Caffau}, {Di Matteo}, {Fabre}, {Fabricius},
  {Frakgoudi}, {Haywood}, {Huckle}, {Hottier}, {Lasne}, {Leclerc},
  {Mastrobuono-Battisti}, {Royer}, {Teyssier}, {Zorec}, {Crifo}, {Jean-Antoine
  Piccolo}, {Turon}, \& {Viala}}]{Katz_etal_2022}
{Katz}, D., {Sartoretti}, P., {Guerrier}, A., {et~al.} 2022, arXiv e-prints,
  arXiv:2206.05902.
\newblock \doarXiv{2206.05902}

\bibitem[{{Kawata} {et~al.}(2018){Kawata}, {Baba}, {Ciuc{\v{a}}}, {Cropper},
  {Grand}, {Hunt}, \& {Seabroke}}]{Kawata_etal_2018}
{Kawata}, D., {Baba}, J., {Ciuc{\v{a}}}, I., {et~al.} 2018,
  Mon.Not.R.Astr.Soc., 479, L108, \dodoi{10.1093/mnrasl/sly107}

\bibitem[{{Khoperskov} {et~al.}(2020){Khoperskov}, {Gerhard}, {Di Matteo},
  {Haywood}, {Katz}, {Khrapov}, {Khoperskov}, \&
  {Arnaboldi}}]{Khoperskov_etal_2021}
{Khoperskov}, S., {Gerhard}, O., {Di Matteo}, P., {et~al.} 2020,
  Astron.Astrophys., 634, L8, \dodoi{10.1051/0004-6361/201936645}

\bibitem[{{Lindegren} {et~al.}(2021){Lindegren}, {Klioner}, {Hern{\'a}ndez},
  {Bombrun}, {Ramos-Lerate}, {Steidelm{\"u}ller}, {Bastian}, {Biermann}, {de
  Torres}, {Gerlach}, {Geyer}, {Hilger}, {Hobbs}, {Lammers}, {McMillan},
  {Stephenson}, {Casta{\~n}eda}, {Davidson}, {Fabricius}, {Gracia-Abril},
  {Portell}, {Rowell}, {Teyssier}, {Torra}, {Bartolom{\'e}}, {Clotet},
  {Garralda}, {Gonz{\'a}lez-Vidal}, {Torra}, {Abbas}, {Altmann}, {Anglada
  Varela}, {Balaguer-N{\'u}{\~n}ez}, {Balog}, {Barache}, {Becciani}, {Bernet},
  {Bertone}, {Bianchi}, {Bouquillon}, {Brown}, {Bucciarelli}, {Busonero},
  {Butkevich}, {Buzzi}, {Cancelliere}, {Carlucci}, {Charlot}, {Cioni},
  {Crosta}, {Crowley}, {del Peloso}, {del Pozo}, {Drimmel}, {Esquej}, {Fienga},
  {Fraile}, {Gai}, {Garcia-Reinaldos}, {Guerra}, {Hambly}, {Hauser},
  {Jan{\ss}en}, {Jordan}, {Kostrzewa-Rutkowska}, {Lattanzi}, {Liao}, {Licata},
  {Lister}, {L{\"o}ffler}, {Marchant}, {Masip}, {Mignard}, {Mints}, {Molina},
  {Mora}, {Morbidelli}, {Murphy}, {Pagani}, {Panuzzo}, {Pe{\~n}alosa Esteller},
  {Poggio}, {Re Fiorentin}, {Riva}, {Sagrist{\`a} Sell{\'e}s}, {Sanchez
  Gimenez}, {Sarasso}, {Sciacca}, {Siddiqui}, {Smart}, {Souami}, {Spagna},
  {Steele}, {Taris}, {Utrilla}, {van Reeven}, \&
  {Vecchiato}}]{Lindegren_etal_2021}
{Lindegren}, L., {Klioner}, S.~A., {Hern{\'a}ndez}, J., {et~al.} 2021,
  Astron.Astrophys., 649, A2, \dodoi{10.1051/0004-6361/202039709}

\bibitem[{{L{\'o}pez-Corredoira}(2014)}]{Lopez-Corredoira_2014}
{L{\'o}pez-Corredoira}, M. 2014, Astron.Astrophys., 563, A128,
  \dodoi{10.1051/0004-6361/201423505}

\bibitem[{{L{\'o}pez-Corredoira} {et~al.}(2018){L{\'o}pez-Corredoira}, {Allende
  Prieto}, {Garz{\'o}n}, {Wang}, {Liu}, \& {Deng}}]{2018martin}
{L{\'o}pez-Corredoira}, M., {Allende Prieto}, C., {Garz{\'o}n}, F., {et~al.}
  2018, \aap, 612, L8, \dodoi{10.1051/0004-6361/201832880}

\bibitem[{{L{\'o}pez-Corredoira} {et~al.}(2020){L{\'o}pez-Corredoira},
  {Garz{\'o}n}, {Wang}, {Sylos Labini}, {Nagy}, {Chrob{\'a}kov{\'a}}, {Chang},
  \& {Villarroel}}]{2020martin}
{L{\'o}pez-Corredoira}, M., {Garz{\'o}n}, F., {Wang}, H.~F., {et~al.} 2020,
  \aap, 634, A66, \dodoi{10.1051/0004-6361/201936711}

\bibitem[{{L{\'o}pez-Corredoira} \& {Sylos
  Labini}(2019)}]{Lopez-Corredoira_Sylos-Labini_2019}
{L{\'o}pez-Corredoira}, M., \& {Sylos Labini}, F. 2019, Astron.Astrophys., 621,
  A48, \dodoi{10.1051/0004-6361/201833849}

\bibitem[{{Lucy}(1974)}]{1974AJ.....79..745L}
{Lucy}, L.~B. 1974, \aj, 79, 745, \dodoi{10.1086/111605}

\bibitem[{{Mr{\'o}z} {et~al.}(2019){Mr{\'o}z}, {Udalski}, {Skowron}, {Skowron},
  {Soszy{\'n}ski}, {Pietrukowicz}, {Szyma{\'n}ski}, {Poleski}, {Koz{\l}owski},
  \& {Ulaczyk}}]{2019ApJ...870L..10M}
{Mr{\'o}z}, P., {Udalski}, A., {Skowron}, D.~M., {et~al.} 2019, \apjl, 870,
  L10, \dodoi{10.3847/2041-8213/aaf73f}

\bibitem[{{Navarro} {et~al.}(1997){Navarro}, {Frenk}, \&
  {White}}]{1997ApJ...490..493N}
{Navarro}, J.~F., {Frenk}, C.~S., \& {White}, S. D.~M. 1997, \apj, 490, 493,
  \dodoi{10.1086/304888}

\bibitem[{{Poggio} {et~al.}(2018){Poggio}, {Drimmel}, {Lattanzi}, {Smart},
  {Spagna}, {Andrae}, {Bailer-Jones}, {Fouesneau}, {Antoja}, {Babusiaux},
  {Evans}, {Figueras}, {Katz}, {Reyl{\'e}}, {Robin}, {Romero-G{\'o}mez}, \&
  {Seabroke}}]{Poggio_etal_2018}
{Poggio}, E., {Drimmel}, R., {Lattanzi}, M.~G., {et~al.} 2018,
  Mon.Not.R.Astr.Soc., 481, L21, \dodoi{10.1093/mnrasl/sly148}

\bibitem[{{Ramos} {et~al.}(2018){Ramos}, {Antoja}, \& {Figueras}}]{Ramos2018}
{Ramos}, P., {Antoja}, T., \& {Figueras}, F. 2018, \aap, 619, A72,
  \dodoi{10.1051/0004-6361/201833494}

\bibitem[{{Recio-Blanco} {et~al.}(2022){Recio-Blanco}, {de Laverny}, {Palicio},
  {Kordopatis}, {{\'A}lvarez}, {Schultheis}, {Contursi}, {Zhao}, {Torralba
  Elipe}, {Ordenovic}, {Manteiga}, {Dafonte}, {Oreshina-Slezak}, {Bijaoui},
  {Fremat}, {Seabroke}, {Pailler}, {Spitoni}, {Poggio}, {Creevey}, {Abreu
  Aramburu}, {Accart}, {Andrae}, {Bailer-Jones}, {Bellas-Velidis}, {Brouillet},
  {Brugaletta}, {Burlacu}, {Carballo}, {Casamiquela}, {Chiavassa}, {Cooper},
  {Dapergolas}, {Delchambre}, {Dharmawardena}, {Drimmel}, {Edvardsson},
  {Fouesneau}, {Garabato}, {Garcia-Lario}, {Garcia-Torres}, {Gavel}, {Gomez},
  {Gonzalez-Santamaria}, {Hatzidimitriou}, {Heiter}, {Jean-Antoine Piccolo},
  {Kontizas}, {Korn}, {Lanzafame}, {Lebreton}, {Le Fustec}, {Licata},
  {Lindstrom}, {Livanou}, {Lobel}, {Lorca}, {Magdaleno Romeo}, {Marocco},
  {Marshall}, {Mary}, {Nicolas}, {Pallas-Quintela}, {Panem}, {Pichon},
  {Riclet}, {Robin}, {Rybizki}, {Santovena}, {Silvelo}, {Smart}, {Sarro},
  {Sordo}, {Soubiran}, {Suvege}, {Ulla}, {Vallenari}, {Zorec}, {Utrilla}, \&
  {Bakker}}]{Recio-Blanco_etal_2022}
{Recio-Blanco}, A., {de Laverny}, P., {Palicio}, P.~A., {et~al.} 2022, arXiv
  e-prints, arXiv:2206.05541.
\newblock \doarXiv{2206.05541}

\bibitem[{{Reid} {et~al.}(2014){Reid}, {Menten}, {Brunthaler}, {Zheng}, {Dame},
  {Xu}, {Wu}, {Zhang}, {Sanna}, {Sato}, {Hachisuka}, {Choi}, {Immer},
  {Moscadelli}, {Rygl}, \& {Bartkiewicz}}]{Reid_etal_2014}
{Reid}, M.~J., {Menten}, K.~M., {Brunthaler}, A., {et~al.} 2014, Astrophys.J.,
  783, 130, \dodoi{10.1088/0004-637X/783/2/130}

\bibitem[{{Romero-G{\'o}mez} {et~al.}(2019){Romero-G{\'o}mez}, {Mateu},
  {Aguilar}, {Figueras}, \& {Castro-Ginard}}]{Gomez2019}
{Romero-G{\'o}mez}, M., {Mateu}, C., {Aguilar}, L., {Figueras}, F., \&
  {Castro-Ginard}, A. 2019, \aap, 627, A150,
  \dodoi{10.1051/0004-6361/201834908}

\bibitem[{{Sartoretti} {et~al.}(2022){Sartoretti}, {Marchal}, {Babusiaux},
  {Jordi}, {Guerrier}, {Panuzzo}, {Katz}, {Seabroke}, {Th{\'e}venin},
  {Cropper}, {Benson}, {Blomme}, {Haigron}, {Smith}, {Baker}, {Chemin},
  {David}, {Dolding}, {Fr{\'e}mat}, {Janssen}, {Jasniewicz}, {Lobel}, {Plum},
  {Samaras}, {Snaith}, {Soubiran}, {Vanel}, {Zwitter}, {Brouillet}, {Caffau},
  {Crifo}, {Fabre}, {Frakgoudi}, {Jean-Antoine Piccolo}, {Huckle}, {Lasne},
  {Leclerc}, {Mastrobuono-Battisti}, {Royer}, {Viala}, \&
  {Zorec}}]{Sartoretti_etal_2022}
{Sartoretti}, P., {Marchal}, O., {Babusiaux}, C., {et~al.} 2022, arXiv
  e-prints, arXiv:2206.05725.
\newblock \doarXiv{2206.05725}

\bibitem[{{Sch{\"o}nrich} {et~al.}(2010){Sch{\"o}nrich}, {Binney}, \&
  {Dehnen}}]{Schonrich_etal_2010}
{Sch{\"o}nrich}, R., {Binney}, J., \& {Dehnen}, W. 2010, Mon.Not.R.Astr.Soc.,
  403, 1829, \dodoi{10.1111/j.1365-2966.2010.16253.x}

\bibitem[{{Sofue}(2020)}]{2020Galax...8...37S}
{Sofue}, Y. 2020, Galaxies, 8, 37, \dodoi{10.3390/galaxies8020037}

\bibitem[{{Wang} {et~al.}(2018){Wang}, {L{\'o}pez-Corredoira}, {Carlin}, \&
  {Deng}}]{2018MNRAS.477.2858W}
{Wang}, H., {L{\'o}pez-Corredoira}, M., {Carlin}, J.~L., \& {Deng}, L. 2018,
  \mnras, 477, 2858, \dodoi{10.1093/mnras/sty739}

\bibitem[{{Wang} {et~al.}(2020){Wang}, {L{\'o}pez-Corredoira}, {Huang},
  {Carlin}, {Chen}, {Wang}, {Chang}, {Zhang}, {Xiang}, {Yuan}, {Sun}, {Li},
  {Yang}, \& {Deng}}]{2020MNRAS.491.2104W}
{Wang}, H.~F., {L{\'o}pez-Corredoira}, M., {Huang}, Y., {et~al.} 2020, \mnras,
  491, 2104, \dodoi{10.1093/mnras/stz3113}

\bibitem[{{Wang} {et~al.}(2022){Wang}, {Hammer}, \&
  {Yang}}]{2022MNRAS.510.2242W}
{Wang}, J., {Hammer}, F., \& {Yang}, Y. 2022, \mnras, 510, 2242,
  \dodoi{10.1093/mnras/stab3258}

\bibitem[{{Widrow} {et~al.}(2012){Widrow}, {Gardner}, {Yanny}, {Dodelson}, \&
  {Chen}}]{2012ApJ...750L..41W}
{Widrow}, L.~M., {Gardner}, S., {Yanny}, B., {Dodelson}, S., \& {Chen}, H.-Y.
  2012, \apjl, 750, L41, \dodoi{10.1088/2041-8205/750/2/L41}

\bibitem[{{Xue} {et~al.}(2008){Xue}, {Rix}, {Zhao}, {Re Fiorentin}, {Naab},
  {Steinmetz}, {van den Bosch}, {Beers}, {Lee}, {Bell}, {Rockosi}, {Yanny},
  {Newberg}, {Wilhelm}, {Kang}, {Smith}, \& {Schneider}}]{2008ApJ...684.1143X}
{Xue}, X.~X., {Rix}, H.~W., {Zhao}, G., {et~al.} 2008, \apj, 684, 1143,
  \dodoi{10.1086/589500}

\end{thebibliography}

\end{document}